\documentclass[12pt,preprint]{aastex}
\newcommand{\Ms}{$M_\odot\,$}                 
\newcommand{\sigHB}{$\sigma_{\rm HB}\,$}                
\newcommand{\vi}{$V\!-\!I\,$}% V-I
\newcommand{\uvv}{$m(1500)\!-\!V\,$}% 15-V

\slugcomment{Accepted: ApJ, 2003, 582 (Jan 1)}

\shortauthors{Sukyoung K. Yi}
\shorttitle{Uncertainties of Synthetic Integrated Colors}

\begin{document}

\title{Uncertainties of Synthetic Integrated Colors as Age Indicators}

\author{Sukyoung K. Yi}
\affil{University of Oxford, Astrophysics, Keble Road, Oxford OX1 3RH, UK and\\
Center for Space Astrophysics, Yonsei University, Seoul 120-749, Republic of Korea\\ yi@astro.ox.ac.uk}

\begin{abstract}
 
        We investigate the uncertainties in the synthetic integrated colors
of {\em simple stellar populations}, 
currently the most popular method of estimating the ages of unresolved 
stellar systems.
        Three types of uncertainties studied here originate
from the stellar models, the population synthesis techniques,
and from the stellar spectral libraries.
        Despite some skepticism, synthetic colors appear to be reliable 
age indicators as long as they are used for select age ranges.
        Rest-frame optical colors are good age indicators
at ages 2 -- 7\,Gyr, mainly due to the clear redward evolution
of hydrogen-burning stars (main-sequence stars and red giants).
        At ages sufficiently large to produce hot horizontal-branch stars, 
the UV-to-optical colors may provide an alternative means for measuring ages.
        This implies that one can use integrated colors as age indicators
for globular clusters in nearby external galaxies and perhaps even for
high redshift galaxies that are passively evolving.
        Such studies may provide important tests of various galaxy
formation scenarios.

\end{abstract}


\section{Introduction}

	Age determination of various stellar populations enhances
our understanding not only on the evolution of stars and galaxies
but also on cosmology.
	Population synthesis has been a very popular tool to do just this.
	For decades, however, the ages of stellar populations inferred by 
population synthesis studies based on the stellar evolution theory
have been a factor of two larger than the age of the universe suggested 
by standard Big Bang cosmology models.
	This is often called ``the age paradox''.
	This has made the efforts to understand the evolution of galaxies 
and cosmology by means of measuring their ages appear hopeless.

	At last, the stellar evolution theory and cosmology seem to 
agree upon the age of the universe.
	Various improvements in input physics have caused significant 
changes in stellar models.
	As a result, the latest isochrone studies --- currently the most 
reliable age dating technique --- now yield 
approximately 20\% smaller ages for Galactic globular clusters than 
previously derived (see Yi et al. 2001).
	At the same time, observational cosmologists found a strong 
evidence of anti-gravity force operating in the universe, which indicates
a substantially larger age for the universe than predicted by the classic, 
matter-dominating flat universe model (Schmidt et al. 1998; 
Perlmutter et al. 1999).
	Although such an agreement does not validate either approach 
by any means, it may imply that both of these fields have matured 
enough that the predictions from stellar astrophysics 
can be compared with observational data with good reliability and perhaps
can even be used to test cosmological models.
	Such predictions are basically temperatures, luminosities, and 
lifetimes at each evolutionary stage, in addition to the overall 
evolutionary pattern.
  	These predictions are being tested in numerous studies
mostly by isochrone users, and the agreement between models and
observed data achieved so far seems quite encouraging (see Yi et al. 2001).

	The isochrone-fitting technique currently cannot reach a
target object beyond the distance to the Magellanic Clouds, however,
because such distant populations are not spatially resolved.
	Thus, astronomers instead construct synthetic models for
integrated properties based on the predictions from stellar astrophysics 
and use them to measure the ages of unresolved populations. 
	Such efforts are made not only on simple systems, such as
star clusters, but also on complex galaxy systems.
	For their capability to reach far in time and space and for
being easy for application, synthetic colors are more popular than ever 
before.
	However, without assessing their reliability (or uncertainty) first,
the level of significance of the analyses is hard to quantify.
	For example, despite the apparent success of the latest isochrone 
fits, stellar astrophysicists still believe that there are many important 
pieces of information that are missing in the stellar model construction 
process.
	Some of these uncertainties may have significant impacts on the
population analyses that are based on synthetic colors.
	Yi et al. (2001) recently showed that their population
synthesis models based on the latest isochrones yielded consistent ages
on Galactic globular clusters to their isochrone-based ages
(see their Figure 18).
	It is the goal of this paper to investigate the impacts of such
uncertainties on such age derivations for various populations.

	In this paper, we investigate three different categories of 
uncertainties.
	In \S 2, we demonstrate the effects of various input parameters
and assumptions that are still poorly understood.
	First, the uncertainties in the stellar models. 
	These are basically the problems related to the solar calibration, 
including the current approximation schemes for convection
and the uniform scaling assumption for the chemical compositions.
	Second are the uncertainties in the synthetic model construction.
	Most notable of these are initial mass function (IMF) and mass loss.
	Last, but not least, are the uncertainties in the stellar 
spectral library.
	In \S 3, we translate such uncertainties into the uncertainties
in the age estimations based on synthetic colors.

	It was Charlot, Worthey, \& Bressan (1996) that first performed a
comprehensive investigation on the uncertainties of the synthetic models.
	They studied the uncertainties that come from using different 
sets of stellar models and stellar spectral libraries.
	In a way, our study is in the same spirit.
	The major difference is that we investigate the effect of
each individual physical assumption separately.
	We then estimate the uncertainties in the final age estimates
stemming from each source of uncertainty.
	This study is only for ``simple'' (single age and metallicity) stellar
populations, and there are further sources of uncertainties in modeling
composite populations, e.g., from star formation history and galactic
chemical evolution.
	Through this study, we hope to provide both synthetic model builders
and users with some guidelines as to what levels of certainties and limits 
they should expect from the models they deal with.

\section{Uncertainties in Input Physics}

\subsection{Uncertainties in the Stellar Models}

	The most fundamental problem in building stellar models is that
there is no strong physical justification for ``the solar calibration''.
	There still are several important pieces of input physics
that need to be understood further in order for us to achieve realistic 
stellar models.
	Most significant of all are convection and chemical evolution.
	Without knowing the underlying physics in such complex mechanisms,
we try to find a solution that works for the sun.
	First, we assume that we know the luminosity, radius, and the age 
of the sun.
	This is already risky, because the solar luminosity measurements may be
uncertain by as much as 1\%.
	Next, we try to match the solar observables with models
by fiddling with poorly-constrained input parameters, such as convection
efficiency and chemical compositions.
	As a result of this step, we have a set of input parameters
that works for the sun.
	We then apply it to all stars that have completely different 
physical and chemical properties.
	The danger of the solar calibration is well known to stellar
astrophysicists.
	In fact, the sun may not even be a typical star.
	In this section, we investigate the sensitivity of synthetic 
population models to uncertainties in the current convection treatments 
and in the heavy element mixture.
	All models shown in this section are based on the assumption of
a Salpeter IMF index (2.35), a mass loss efficiency $\eta = 0.65$ 
and a dispersion $\sigma = 0.02$\Ms, and the Lejeune et al.'s revised Kurucz 
spectral library (see \S 2.2 and 2.3).

\subsubsection{Mixing Length Approximation (MLA)}

	Standard stellar models are still based on the half-a-century-old 
Mixing Length Approximation (MLA) for describing the convective energy 
transport.
	Within the framework of MLA, we assume a fixed value of
the ratio of mixing length to pressure scale height, $l/H_{p}$ (also known 
as ``the mixing length parameter'', $\alpha_{\rm MLA}$) that directs 
convection in the same manner in all different environments.
	The mixing length parameter primarily governs the stellar radius,
and we find this value from the solar calibration.
	Obviously, there is a problem in this approach.

	MLA is not a forward physical description, but a parameterisation.
	Thus, even if it can be justified in sun-like stars, 
it may break down in environments that are different from that of the sun.
	Such environments may include not only chemically-different stars
but also different evolutionary stages of the same star.
	Although there are efforts to find more realistic descriptions of 
convection than MLA (e.g., Nordlund \& Dravins 1990; Canuto 1990; 
Kim et al. 1996; Freytag, Ludwig, \& Steffen 1966), 
there has not yet been a release of a comprehensive set of stellar models 
based on a more mature convection prescription.
	Therefore, it is important to understand the acceptable range of the
mixing length parameter within the framework of MLA and its impact on
the reliability of synthetic population models.

	Figure 1 shows the effects of the mixing length parameter 
to the shape of isochrones at two ages. 
	A larger value of $\alpha_{\rm MLA}$ causes a more efficient energy 
transport and thus the stellar models to appear hotter (bluer).
	The larger value (1.74) is currently the favoured value of the 
Yale-Yonsei collaboration (Yi et al. 2001) for matching the solar properties.
	Currently, the uncertainties in other input parameters
(e.g., metal-to-hydrogen ratio measured in the sun, and solar helium abundance)
prevent an accurate determination of $\alpha_{\rm MLA}$.
	However, all latest calibrations are clustered around
$\alpha_{\rm MLA} \approx 1.5$ -- 2.0 (e.g., Girardi et al. 2000; 
Vandenberg et al. 2000).
	Thus, our investigation with two values (1.74 and 1.5) 
well-represent a mean and one extreme case.

	The effects of $\alpha_{\rm MLA}$, which is not sensitive to age, 
naturally translates into uncertainties in the spectral energy distribution, 
as shown in Figure 2.
	The fluxes are normalized to the $V$-band flux of our standard
model, i.e., that of $\alpha_{\rm MLA}$=1.74.
	The synthetic spectral energy distribution with a larger value 
of $\alpha_{\rm MLA}$ appears hotter.
	Figure 3 shows the color evolution of a sample population of
a typical metallicity of Galactic globular clusters, i.e., [Fe/H]=$-$1.3.
	The $m(1500)-V$ is defined as 
\begin{equation}
m(1500)-V = -2.5~log~\frac{<\!f_{1500}\!>}{<\!f_V\!>}
\end{equation}
where $<\!f_{1500}\!>$ and $<\!f_{V}\!>$ are defined by 
averaging the flux within the ranges 1250 $-$ 1850 \AA\, and
and 5055 $-$ 5945~\AA, respectively.

	The difference in optical and near-infrared colors ($UBVRI$) 
seems insignificant at first sight.
	However, the color evolution with time is so small that even 
such small uncertainties in synthetic colors can translate into substantial
uncertainties in the age estimate.
	For example, when \vi is used, the uncertainty in the mixing
length parameter alone can cause up to a 25\% uncertainty in the age 
estimate.

	It should also be noted that for low metallicities and for the
age range shown here, \ub is not a good age indicator at all, because
it does not behave linearly with decreasing main-sequence (MS) turn-off (MSTO)
temperature (i.e., increasing age) under such conditions.

\subsubsection{Convective Core Overshoot}

	Another important source of uncertainty in convection is the amount
of convective core overshoot (hereafter, OS).
        Convective core overshoot, the importance of which was first 
pointed out by Shaviv \& Salpeter (1973), is the inertia-induced penetrative 
motion of convective cells, reaching beyond the convective core as defined 
by the classic Schwarzschild criterion. 
        Stars develop convective cores if their masses are larger than
approximately 1 -- 2 \Ms, typical for the MSTO stars in 1 -- 5 Gyr-old 
populations, depending on their chemical compositions.
        Since the advent of the OPAL opacities, various studies have 
suggested a modest amount of OS; that is, OS $\approx 0.2 H_{p}$, where 
$H_{p}$ is pressure scale height (see references in Yi et al. 2001).
        Thus, we have adopted OS=0.2 in our standard models.
        OS has many effects on stellar evolution.
        Among the most notable are its effects on the shape of the MSTO and 
on the ratio of lifetimes spent in the core hydrogen burning stage (i.e., 
the MS phase) 
and in the shell hydrogen burning stage (the red giant branch: RGB).
        The impacts of such effects on the isochrone and on the integrated
spectrum have been discussed by Yi et al. (2000).

	OS is by definition effective only for the stars massive enough to
develop a convective core, so its effects are visible only in young
populations.
	Figures 4 -- 6 show that the effects are clearly visible in 1\,Gyr 
models.
	But they quickly disappear already at 2\,Gyr as soon as MSTO 
stars begin to have a radiative core.
	Accurate borders depend on the metallicity in question,
as listed in Table 2 of Yi et al. (2001).
	Young populations with the effects of OS display bluer spectra.
	The effects of OS may appear negligible but are important in the
age estimation of such young populations.
	In particular, Yi et al. (2000) demonstrated that 
precise age derivations for high-$z$ galaxies,
which is a powerful method for constraining the galaxy formation epoch
(Dunlop et al. 1996), are heavily subject to this uncertainty.

\subsubsection{Chemical Composition}

	The chemical composition has many important impacts on stellar 
evolution.
	The first is on the nuclear energy generation.
	For example, the CNO cycle uses CNO elements as catalysts, and thus
their higher abundance causes a more effective energy generation if 
the right conditions are met.
	Another effect is on the opacity.
	Atoms with many possible transitions (most notably Fe) are 
particularly important, because a small increase in their abundance can 
raise opacity significantly.
	Increase in opacity obviously causes the spectrum to appear redder,
but it also governs the energy transport and energy generation.

	The effects of helium abundance are not to be ignored, either.
	In the stellar cores in most evolutionary stages, helium abundance
is the key factor in opacity and determines the internal temperature profile, 
which controls energy generation rates.
	For this reason, helium abundance primarily determines the pace of 
stellar evolution.
	However, this effect is strong only at metallicities above
solar.
	Since this study is more focused on low metallicities (approximately
solar and below solar), the uncertainty caused by the helium abundance
is not explored in this paper.

	Despite these potential complexities, the concept of metallicity 
often seems to be adopted too naively.
	Metallicity is generally given only in $Z$ or in [Fe/H], 
but departures in some elements from this mean quantity (usually solar)
cause important changes in stellar evolution.
	Most notable is perhaps the $\alpha$-element enhancement,
which is ubiquitous in metal-poor stars and perhaps in super-metal-rich 
environments as well.

	It is a common misconception that $\alpha$-enhancement 
changes the stellar evolution in a very mysterious way.
	Figure 7 shows two isochrones with and without $\alpha$-enhancement 
but for the same total $Z$.
	We are basically comparing two models of the same total metallicity 
in total metal abundance $Z$ but of different $\alpha$-element abundances.
	In other words, the $\alpha$-enhanced models have higher abundances in 
$\alpha$-elements but lower abundances in other heavy elements.
	This figure shows no detectable difference.
	This implies that $\alpha$-enhanced models can be mimicked 
by more metal-rich standard (non-$\alpha$-enhanced) models after an 
appropriate scaling, such as the one suggested by Salaris, Chieffi, \& 
Straniero (1993).
	So, when we compare two metal-poor models with the same total 
metal abundance, $\alpha$-enhancement does not cause a significant 
difference in stellar evolution models.
	Naturally, little difference is expected in the color evolution 
models and in the age estimates based on them, either.
	The effect of $\alpha$-enhancement is somewhat more conspicuous
when metallicity is equal or higher than a solar value; $\alpha$-enhanced 
stellar models are somewhat bluer.
	But, the magnitude of the difference is still small.

\subsection{Uncertainties in the Synthetic Model Construction}

	The synthetic population models shown in the previous section
are all based on popular choices of the input parameters used in the
population synthesis; that is, the Salpeter initial mass function,
and the mass loss parameters that reproduce the 
horizontal-branch (HB) morphology of Galactic globular clusters.
	However, whether such choices are valid in all populations
is debatable.
	In this section, we investigate the contribution of the population
synthesis input parameters to the uncertainty of synthetic colors.
	All models shown in this section are based on the assumption of
$\alpha_{\rm MLA}$=1.74, OS=0.2, and [$\alpha$/Fe]=0 (see \S 2.1).

\subsubsection{Initial Mass Function (IMF)}

	Since its first introduction by Salpeter (1955), the Salpeter 
IMF has been the most popular ingredient in galaxy population synthesis
studies.
	However, whether the same slope derived from the local population 
would be valid for describing different populations formed in different
environments is debatable. 
	Besides, the departure of the IMF from the power law below 
approximately a solar mass is no longer beyond any doubt
(e.g., Miller \& Scalo 1979; Kroupa, Tout, \& Gilmore 1990).
	A single power-law with one slope is not an accurate
representation of the true IMF, and it may lead to seriously inaccurate
synthetic models.
	Thus, we investigate the effects of IMF by choosing three 
significantly different IMF slopes in the context of the power law:
1.35, 2.35 (Salpeter), and 3.35.

	The low-mass departure is not a serious issue in the integrated
luminosity evolution of a population, because such low-mass stars
are in the faint, MS stage during the most of the age range that we are
interested in. 
%	Even if we choose a significantly different slope, such as
%the one suggested by Kroupa et al. (1990), x(IMF)=0.7, as a much better 
%representation than the Salpeter slope for the low-mass mass function,
%one would not see much of its effect once a population is older than
%a few billion years.
	For the age range that we are studying in this paper --- i.e.,
1 through 15\,Gyr (or, a MSTO mass range of 2 -- 0.9\Ms), the Salpeter slope 
(2.35) appears to be a good approximation (Sagar, Munari, \& de Boer 2001).
	Thus, the range of slope we have chosen seems sufficiently large
to represent the uncertainty of the IMF slope.

	Figure 8 shows the effect of the IMF slope to the integrated spectra.
	The comparison is shown for two ages (3 and 10\,Gyr).
	Two dotted lines are the models for the smaller IMF slope.
	Younger models are normalized to the $V$-band flux of their older
(10\,Gyr) model.
	Two old models with different IMF slopes are on top of another,
being indistinguishable.
	The difference in luminosity evolution is notable.
	When the IMF slope is uncertain by 1.0, as shown here,
the luminosity evolution model is uncertain by 0.5 mag between the two
epochs of 3\,Gyr and 10\,Gyr of ages; figuratively speaking, between 
redshift 1 and 0 (assuming a $\Lambda$-dominated flat universe).
	This difference in the luminosity evolution, 0.5 mag, is approximately
half of the luminosity evolution observed in elliptical galaxies that are
often believed to have been passively evolving.
	Thus, it is extremely important to learn about the true IMF
in order to interpret the observed luminosity evolution correctly.
	If the true IMF slopes for old populations 
(formed at high redshifts) were different from that of the local field
stellar sample (such as the Salpeter slope), such simplistic approaches
in interpreting luminosity evolution data are bound to be misleading.

	The impact of the IMF slope also appears in the color evolution, 
although less severely.
	Figure 9 shows this.
	Synthetic models become redder as a smaller IMF slope is used,
because a smaller slope means a larger number density of more massive 
MS stars, which evolve faster to become red giants.

\subsubsection{Mass Loss Efficiency}

	Another key area of uncertainty is the treatment of mass loss on the 
giant branch, which predominantly governs the temperatures of post-giant 
branch stars.
	Post-giant branch stars account for 30\% of the total light in 
the $V$-band at a 2\,Gyr age,  and 40\% in the near UV at 12\,Gyr;
and, thus, outdated values for mass loss lead to serious errors in the 
flux distribution. 

	Whether the Reimers (1975) empirical formula is valid for the mass loss
in all stars is debated (see references in Willson 2000).
	Most of the empirical mass loss measurements are conducted on
high-mass (above solar) asymptotic giant branch stars, and it may
not accurately describe the mass loss that takes place on the
RGB of low-mass stars.
	Yet, it is still the most popular prescription in population
synthesis studies.
	Reimers' empirical formula is as follows:
\begin{equation}
\frac{d M}{d t} = -4\,\times\,10^{-13}\,\eta\,\frac{L}{g\,R},
\end{equation}
where $L$ is the luminosity, $g$ is the surface gravity, and $R$ is the radius.

	Even when we assume that Reimers' formula provides a good 
approximation, its fitting parameter $\eta$ is still somewhat uncertain.
	Renzini (1981), among others, noted that one can determine $\eta$
in low-mass stars by comparing the mean mass of the red giants with that of 
HB stars in globular clusters.
	The mean mass of red giants can be relatively well-derived from the
MSTO isochrone fitting, while that of HB stars can also be determined
through pulsation analysis or synthetic HB construction, albeit with
a somewhat lower precision.
	Renzini (1981) found $\eta$ of 0.3 based on the information 
available then, while the more recent work of Yi et al. (1997; 1999) found a 
factor of two larger efficiency favoured.
	It is, however, very difficult to compare one value to another
because it depends not only on the stellar evolutionary tracks that were 
used to compute the mass loss but also on the ages assumed for the globular
clusters in question.
	Thus, the actual values of the efficiency parameter or 
of the mass loss derived from different studies may be slightly different 
from one study to another.
	However, they must be able to reproduce the observed HB morphology
using the mass loss efficiency they have chosen.
	Yi et al. (1999)'s choice of $\eta$ focused on the data of Galactic 
globular clusters of typical metallicity, i.e., [Fe/H]=$-$1.3.
	However, Yi et al. (1997) also remarked about a possible positive
correlation between the mass loss efficiency and metallicity, which is
consistent with the theoretical suggestion of Willson (2000). 
	It seems that an accurate computation of mass loss as a function 
of stellar mass (time) and of metallicity is still a difficult task.

	The mass loss efficiency $\eta$ is appears to be uncertain
at least by a factor of two to three --- approximately in the range of 
0.3 through 1.0.
	However, this uncertainty is mostly due to its ill-understood
sensitively to metallicity.
	For a given metallicity, the synthetic color-magnitude diagram 
(CMD) fitting technique
suggests approximately 20\% uncertainty in $\eta$.
	Therefore, we investigate the outcome by the use of three
values of mass loss efficiency: 0.55, 0.65 (standard), and 0.75.
	Figure 10 shows the result.

	When a larger value of $\eta$ is used, stars lose more mass
on the RGB and become lower-mass, hotter, HB stars.
	This results in a higher UV flux.
	Old Galactic globular clusters in general exhibit UV spectra 
as seen in the model with $\eta = 0.65$ (curve in the middle in the UV).
	Most clusters are bracketed by the other two models in terms
of UV strength (see Yi et al. 1997). 
	This effect becomes noticeable only after the population 
is old enough to develop a blue HB.
	We illustrate this in Figure 10 by showing models at two ages.
	Figure 11 shows this in color evolution.
	The non-monotonicity in \bv at ages 9 -- 15\,Gyr is due to 
the gradual development of a bluer HB with increasing age.
	The blueward turn-around at age 10\,Gyr is due to the
first appearance of the blue HB.
	The redward turn a few Gyr later is due to two effects.
	First, at such large ages, low-mass stars evolve very slowly and 
thus a smaller number of stars leave the hydrogen-burning stage to enter 
the HB phase.
	Second, some of red giants, which are already quite low in mass, lose
too much mass during this long evolution on the RGB and consequently
lack enough mass in the envelope to ignite the helium core flash.
	These stars may skip the HB phase and become faint planetary nebulae.

	The color evolution beyond 13 -- 14\,Gyr is, however, quite uncertain
because there are complex interplays between stellar evolutionary pace,
the growth of core, mass loss, and the helium core flash.
	It should also be noted that similar effects from the HB morphology
is believed to be important in spectral line index evolution as well 
(Lee, Yoon, \& Lee 2000).

\subsubsection{Mass Loss Dispersion}

	The dispersion in mass loss, evident from the color spread on the HB,
has a similar, albeit smaller, effect, as Yi et al. (1997) demonstrated.
        The over-simplistic clump-star assumption for all post-giant phases 
is still routinely employed, inevitably resulting in inaccurate models.
	This dispersion, often denoted by \sigHB (in \Ms), is probably 
caused by the dispersion in mass loss on the RGB.
	It can be estimated by comparing synthetic HBs to observed ones.
	Various studies have adopted the single Gaussian mass function 
to construct synthetic HBs, and the value of the Gaussian mass 
dispersion has been reported to be around 0.02 \Ms with some 
uncertainty (Lee, Dmarque, \& Zinn 1994; Catelan 2000). 
	Except for some clusters that show a strong multi-modal color
distribution on the HB (e.g., M\,15), no Galactic globular cluster 
shows a mass dispersion considerably larger than twice of this value.
	Thus, we chose three values of \sigHB, 0.001, 0.02, and 0.04,
representing a single-peaked clump, normal, and wide mass dispersion
assumptions.

	Figures 12 -- 13 show the results.
	The effects are qualitatively similar to those of $\eta$, although
somewhat smaller.
	They appear only at large ages, where a population develops an HB
that is sufficiently blue.

\subsection{Uncertainties in the Stellar Spectral Libraries}

	Another significant source of uncertainty is the quality 
of the spectral libraries that are used to map theoretical stars into 
the observational domain.
	For transforming theoretical quantities to colors, 
color transformation tables are more popular than spectral libraries.
	Among the most popular color transformation tables are
that of Green, Demarque, \& King (1987) and that of Kurucz (1979; 1992).
	The Green et al. table is a semi-empirical one in the sense
that they modified the old Kurucz table (1979) using empirical stellar
data.
	As a result, the Revised Yale Isochrones based on the Green
et al. table matched the observed CMDs of star clusters reasonably well.
	More than a decade later, Kurucz made a major update on his
stellar spectral library and a corresponding color transformation table
(Kurucz 1992).
	This new, purely theoretical library is still reported to seriously 
mismatch the cool-star data (e.g., Morossi et al. 1993), however.
	An effort to construct a semi-empirical spectral library
based on the latest Kurucz library was made by Lejeune and collaborators
(Lejeune, Cuisinier, \& Buser 1997; 1998).
	Yi et al. (2001) find that this Lejeune et al.-calibrated Kurucz
library (hereafter, the LCBK library) is qualitatively as good as 
that of Green et al. in reproducing observed CMDs.
	The LCBK library appears to be better calibrated in high 
metallicities and on the RGB, while the Green et al. table performs 
better for low metallicities and on the MS.
 	Consequently, Yi et al. (2001) have released their newly computed 
isochrones using both color transformation schemes.

	If we grant that we cannot yet clearly prefer one of these libraries, 
we are immediately trapped by the fact that they yield substantially 
different color evolution models.
	Figures 14 and 15 show the difference in color for an approximately
solar metallicity stars but for two different evolutionary stages; 
MS stars and red giants.
	Compared to the libraries are empirical stellar data (plus symbols), 
kindly provided by Worthey (priv. comm.).
	Figure 14 shows that the LCBK table is correcting the dwarf data of
the Kurucz table substantially in the low-temperature regions.
	As a result, the corrected table is very close to that of Green et al.
(GDK).
	In $U$$-$$B$ and $B$$-$$V$, the GDK colors seem to match the empirical
data better.
	The difference between the LCBK table and the GDK table
seems small but large enough to stand out in the MS isochrone fitting.
	However, it is the comparison of low-gravity stars (e.g. red giants)
that is striking.
	Figure 15 shows that the difference between different color
calibrations is very large in certain temperature ranges.
	Similar differences exist in other metallicities.
	It should be noted that the empirical calibration is still 
quite poor, espeicially at
super-solar metallicities because stellar sample are scarce; 
and for this reason, modeling super-metal-rich systems (e.g. giant 
elliptical galaxies) is not trivial.
	Figures 15 may give readers an impression that the LCBK colors
match the empirical data very well.
	However, the truth is that the empirical temperature-color relations
(shown as plus symbols) suffer from large ($\lesssim 1$~mag in $U$$-$$B$) 
uncertainties, as shown evidently in Figure 4 of Lejeune et al. (1997).
	Obviously, such differences will appear in the synthetic spectra
and colors as uncertainties.

	Figure 16 shows a comparison of the synthetic model spectra based on
two different spectral libraries; the Kurucz library and the LCBK library.
	The Green et al. calibration was made only on colors, but not
on the spectral library, and thus it is not available for spectral synthesis.
	But, the difference between the LCBK library and the Kurucz
library is large enough to represent the current level of uncertainty 
in the spectral library.
	The model based on the LCBK library is in general redder for
a given age.
	This appears systematically in \bv and \vi, as shown in Figure 17.

\begin{table*}
 \centering
 \begin{minipage}{140mm}
  \caption{Sources of uncertainties in population models and their impacts.}
  \begin{tabular}{@{}ccc@{}}
\hline
Uncertainty                 & Range studied (standard in brackets) & Ages of large impact\\
\hline
Mixing Length Approximation & $\alpha_{MLA}$=1.5 \& (1.74) & all ages\\
Convective Core Overshoot   & OS=(0.2) \& 0.0 $H_{P}$      & age$\la$2\,Gyr\\
$\alpha$-enhancement        & [$\alpha$/Fe]=(0.0) \& 0.3   & none if $Z$ is fixed\\
Initial Mass Function Slope & x=1.35, (2.35), \& 3.35      & age$\la$4\,Gyr\\
Mass Loss Efficiency        & $\eta$=0.55, (0.65), \& 0.75 & age$\ga$8\,Gyr\\
Mass Loss Dispersion        & $\sigma_{HB}$=0.01, (0.02), \& 0.04 & age$\ga$8\,Gyr\\
Spectral Library            & Kurucz (1992), \& Lejeune et al. (1997, 1998) & all ages\\
\hline
\end{tabular}
\end{minipage}
\end{table*}

\section{Uncertainty in Synthetic Colors}

	The uncertainties in synthetic colors caused by the uncertainties
in various input physical assumptions are shown in Figure 18 without legends.
	Basically, the mixing length parameter, the IMF slope and the 
spectral library make moderately important contributions to the 
uncertainty over all ages.
	On the other hand, convective core overshoot is important
only at small ages (less than approximately 2\,Gyr), while
the mass loss efficiency and the mass dispersion are important only 
at large ages.
	If total metallicity is kept the same, $\alpha$-enhancement
plays little role.
	This is summarised in Table 1.

	The total uncertainty (standard deviation) is shown in Figure 19
as inner (smaller) error bars.
	In the worst scenario, if all sources of uncertainties 
behave in the same direction, the maximum uncertainty in colors
as a function of age would appear as the outer (larger) error bars.
	Such a pessimistic outcome is very unlikely, however.
	The errors in the \uvv color at ages smaller than 10\,Gyr are
underestimated.
	This is because we are not exploring the impact of the 
post asymptotic giant branch (PAGB) stellar mass -- the prime factor 
that determines the UV colors at small ages -- as a source of uncertainty, 
which is poorly understood.
	At large ages, HB stars are likely to dominate the UV flux.
	Thus the uncertainty regarding the PAGB mass is probably negligible.

	It is clear that \ub is not a good age indicator at all 
at least for this metallicity, as it behaves non-monotonically with 
respect to age.
	The non-monotonicity disappears in metal-rich models,
but we feel that \ub is still poorly calibrated relative to other optical
colors.
	Both \bv and \vi seem reasonable as age indicators, but only 
at small ages.
	Once the population is older than approximately 7 -- 8\,Gyr,
the spectral evolution in the visible bands is minimal.
	Besides, at somewhat larger ages, populations gradually develop
a larger number of blue HB stars that counteract the redward evolution 
of the MSTO, causing non-monotonicity.
	Thus, we suggest that \bv and \vi should not be used to derive 
the ages of  systems older than 7 -- 8\,Gyr.
	At larger ages than that, \uvv becomes our choice.

	The estimated errors of age estimates (one sigma) in the range of 
ages 2 -- 7\,Gyr, 
as shown in Figure 19, are typically 0.01 -- 0.02 mag both in \bv and \vi.
	For ages 1 -- 2\,Gyr, the uncertainty is dominated by the 
uncertainty in the overshoot parameter and can be as large as 0.1\,mag.
	In this sense, the best results of age determination can be 
achieved when synthetic models are applied to 2 -- 7\,Gyr old populations.

\section{Uncertainty in Age Estimates Derived using Synthetic Colors}

	The uncertainties in synthetic colors directly translate into
uncertainties in age estimates.
	Figure 20 and Table 2 show the level of uncertainty in age 
estimates for two metallicities.
	In Figure 20-(a), we show only the metal-poor model, because
the UV spectral evolution of metal-rich populations have not been 
calibrated sufficiently, in particular in terms of mass loss efficiency.
	This is because there are not enough old, metal-rich 
{\em simple} stellar populations that can be used to constrain mass loss 
efficiency at high metallicities. 
	Even in \bv and \vi, shown in Figure 20-(b) and (c), 
we do not have the same confidence in our old, metal-rich models as we do
in metal-poor models for this reason.

	The uncertainty is smallest at ages 2 -- 5\,Gyr.
	If a population has formed at a high redshift (say at 5), 
it would be this old at redshift 1 -- 2.
	This demonstrates a clear advantage of studying redshift 1 -- 2
objects.
	At $z$=1, rest-frame \bv would be approximately $I\!-\!J$.
	Any color similar to this would work as a useful age indicator
for old populations.

	Most of the detectable distant populations are currently metal-rich.
	Synthetic colors are powerful probes of their ages, but
only when they are in the age range of 2 -- 5\,Gyr.
	Thus, we call for caution in all studies that aim to 
derive the galaxy evolution history by means of population synthesis 
models.
	Rest-frame \bv is affected by the uncertainty in mass loss
and modeling \vi is seriously hampered by the uncertainties in the
spectral library.

\begin{table*}
 \centering
 \begin{minipage}{90mm}
  \caption{Uncertainty in Age Estimates (Gyr) based on Integrated Colors.}
  \begin{tabular}{@{}rrrrrrrr@{}}
\hline
Age     & \multicolumn{3}{c}{[Fe/H]=$-$1.3} & \multicolumn{4}{c}{[Fe/H]=0.0}\\
(Gyr)   & \uvv & \bv & \vi & & \bv & \vi & \\
\hline
2	&  --        & 0.4 & 0.5 & & 0.4 & 0.3  &\\
4	&  --	     & 0.2 & 0.4 & &  0.5 & 1.2 & \\
7	&  --  	     & 0.7 & 0.9 & &  0.7 & 2.0 & \\
11	&  0.5       & --  & --  & &  --  & --   &\\
15	&  0.5       & --  & --  & &  --  & --  & \\
\hline
\end{tabular}
\end{minipage}
\end{table*}

	It is already very difficult to construct reliable, metal-rich
simple stellar population models (single metallicity, coeval populations) 
in $UBV$ bands.
	To make the matters worse, galaxies are not simple stellar 
populations at all.
	Population analyses on galaxies therefore should be performed with 
much caution.

	Despite our cautionary remarks, it is encouraging to find 
that the error bars in age estimates are so small.
	Our analysis suggests that the level of reliability of the current
population synthesis models is likely to allow us to distinguish
13 Gyr-old populations from 8 Gyr-old ones and 8 Gyr-old one from
2 Gyr-old ones with at least 2 -- 3 sigma confidence.
	Therefore, it is possible to use such broad band colors
to determine the ages of simple stellar populations, such as 
globular clusters, in external galaxies.
	Precise measurements of the ages of globualr clusters in 
elliptical galaxies have been of particular interest, because they 
test elliptical galaxy formation scenarios, which is a key issue
in modern cosmology.

	Because the UV to optical color is a good age
indicator for old populations, one can apply a two-color diagram,
such as the one shown in Figure 21, to observed data in order to
distinguish old populations from young ones.
	Once again, we include only metal-poor models in this
plot because we are less confident about the UV spectral evolution of
metal-rich populations.
 	If observed data are clustered only along one narrow straight 
sequence as shown on the left of the shaded box in Figure 21,
the data would imply that the populations are all substantially younger
than Milky Way.
	On the other hand, outliers toward the right-hand side of this
diagonal sequence, i.e., in the shaded box, would be considered older 
populations.
	Overplotted are Galactic globular cluster data.
	The UV-to-V colors and $B$-$V$ are from Dorman, O'Connell, \& Rood 
(1995) and Harris (1996), respectively.
	All Galactic globular clusters are populating in the shaded box,
which is consistent with their large ages.
	This and similar analyses may be useful to break the infamous
age-metallicity degeneracy (Worthey 1994) and derive the ages of 
simple stellar populations, such as external star clusters.

\section{Conclusions}

	Despite some skepticism, synthetic integrated colors 
appear to be reliable age indicators in select age ranges.
	Rest-frame optical colors are good age indicators
at ages 2 -- 7\,Gyr, mainly due to the clear redward evolution
of core hydrogen-burning stars.
	At larger ages optical colors fail to be reliable age 
indicators for two reasons.
	First, at large ages spectral evolution is very slow.
	Second, as MS stars become less massive with 
increasing age, a population develops hotter helium-burning stars, which
causes non-monotonic evolution of optical colors.
	At sufficiently large ages, UV-to-optical colors may
provide an alternative means of measuring ages.
	This can be applied to measure the ages of the globular clusters
in nearby bright elliptical galaxies.
	Such studies may provide important tests against various galaxy
formation scenarios.

	We performed this investigation mainly on metal-poor populations,
because we feel considerably less confident about input assumptions in
modeling for metal-rich populations.
	The most serious problem in modeling metal-rich systems is the
uncertainty in mass loss in low-mass, metal-rich stars.
	We believe that the most reliable way to constrain mass loss
for such systems is to match their HBs with synthetic HB models.
	Thus, it is essential to collect accurate CMDs of metal-rich 
simple stellar populations.

	The caveat of this analysis is that the estimated uncertainty
in age derivation may not be realistic to populations of different 
metallicities and to composite populations.
	However, analyses similar to the one shown in Fig. 21 may be 
able to constrain the ages of simple stellar popuations regardless of 
their metallicities.

	The impacts of the various uncertainties studied here have been
investigated by means of the Yale stellar evolution code and the
Yi's population synthesis code.
	The magnitudes of such impacts could vary when different codes
are used, for instance due to the use of different nemurical techniques, 
but probably only slightly.
	Thus, any disagreement between population models substantially larger
than that demonstrated in this study would be difficult to justify.

\section*{Acknowledgments}

I am grateful to the referee, Guy Worthey, for constructive 
comments, stellar color data used in Figures 14 and 15, and 
for an insightful suggestion regarding Figure 21.
I thank Thibault Lejeune, Jo\~ao Fernandes, Steve Zepf, and many others 
who encouraged me to start this investigation.
Part of this work was performed while I was at California Institute of
Technology. This work was supported by the Creative Research Initiative 
Program of the Korean Ministry of Science \& Technology (grant).

\epsscale{0.8}
\begin{figure}
\plotone{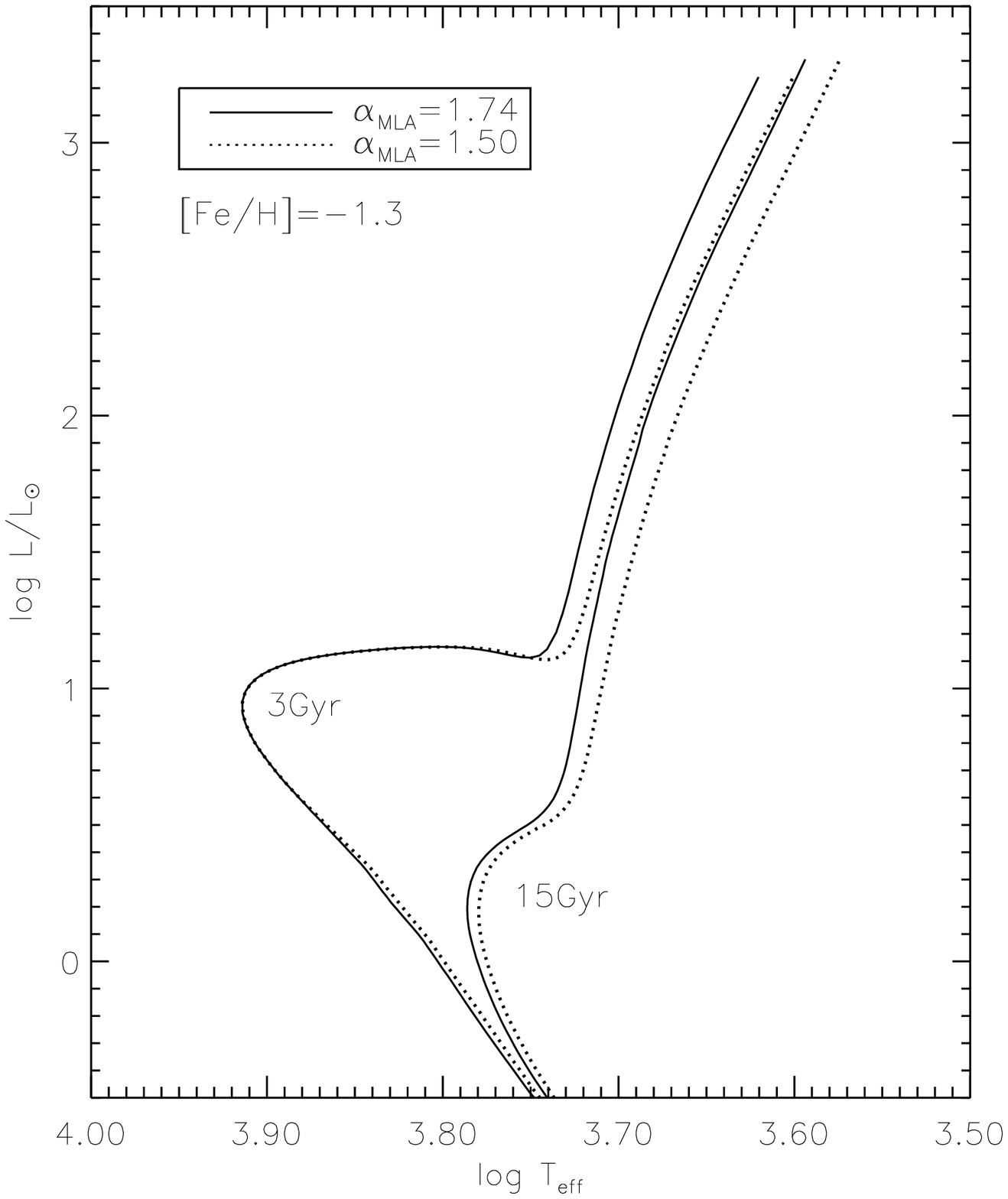}
\caption{The effects of mixing length parameter ($\alpha_{\rm MLA}$) to the 
shape of isochrones. A larger value of $\alpha_{\rm MLA}$ causes the 
stellar models to appear bluer. 
In order to single out this effect, all other input parameters are kept 
the same as in our standard model (see Table 1).}
\end{figure}

\begin{figure}
\plotone{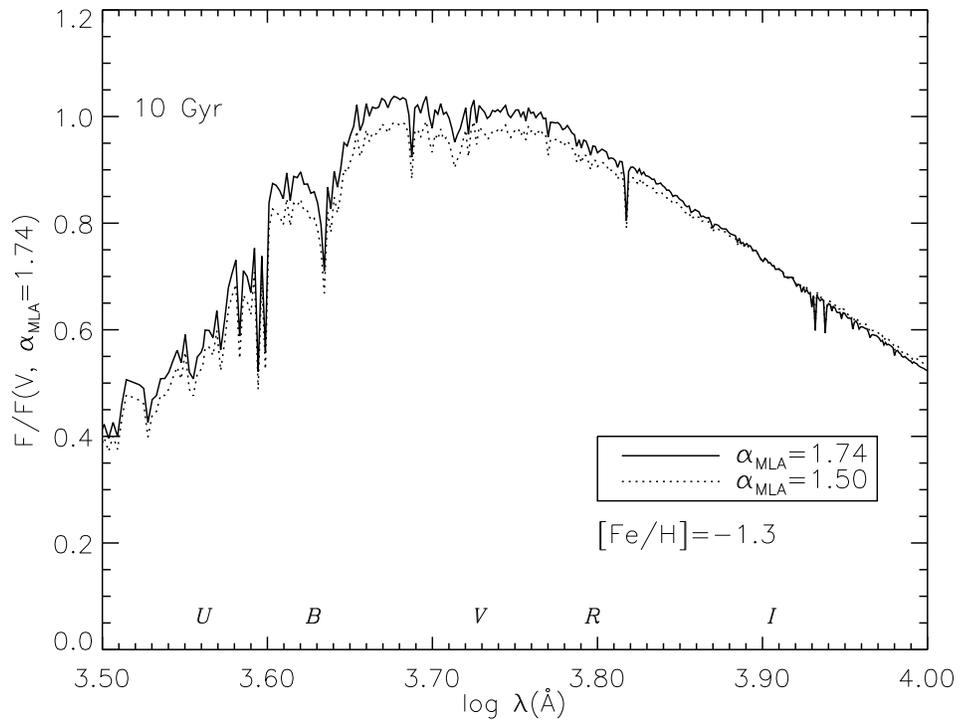}
\caption{The effects of mixing length parameter ($\alpha_{\rm MLA}$) to the 
integrated spectrum. A larger value of $\alpha_{\rm MLA}$ causes the 
population models to appear bluer.}
\end{figure}

\begin{figure}
\plotone{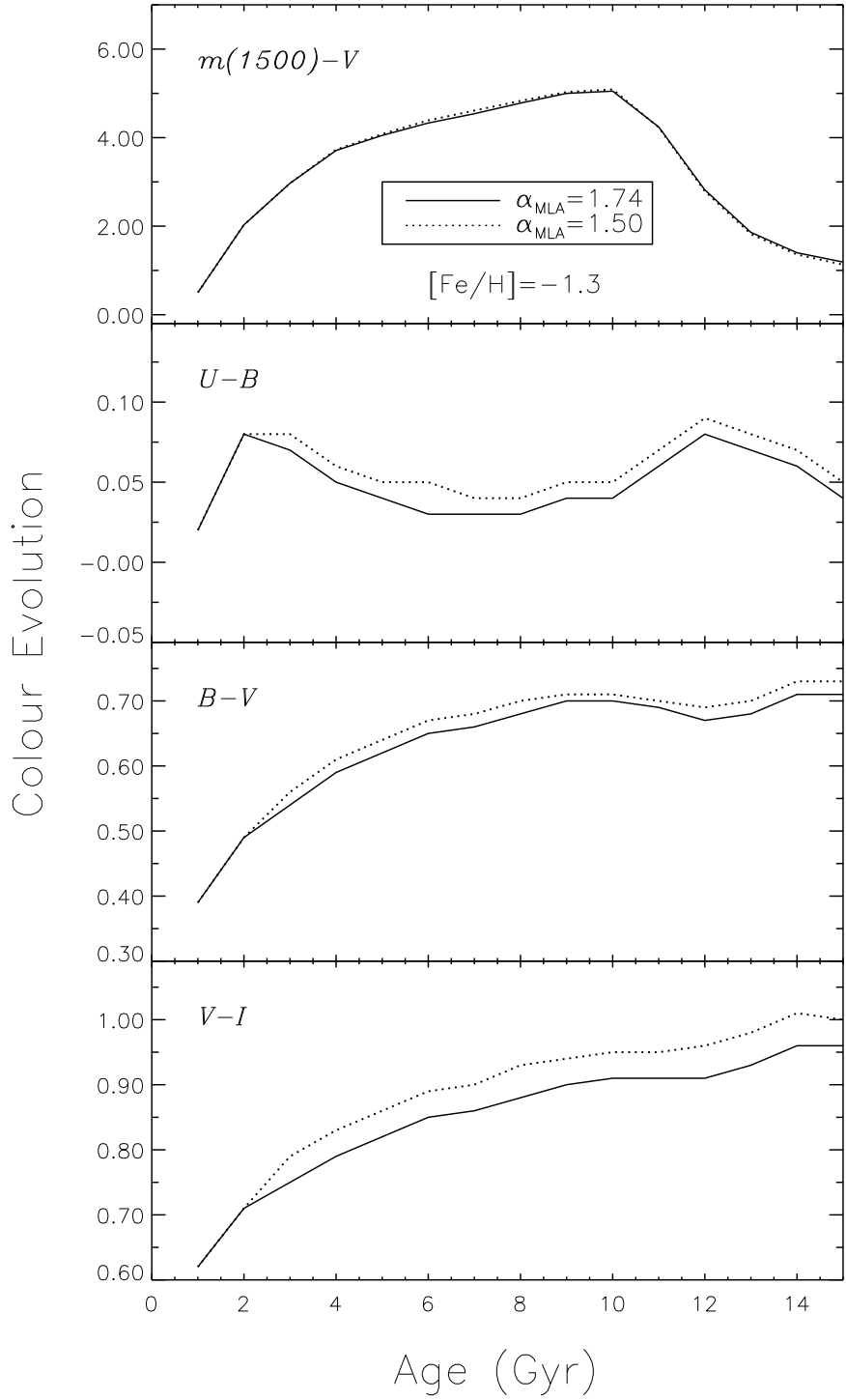}
\caption{The effects of the mixing length parameter ($\alpha_{\rm MLA}$) to 
the synthetic integrated colors.}
\end{figure}

\begin{figure}
\plotone{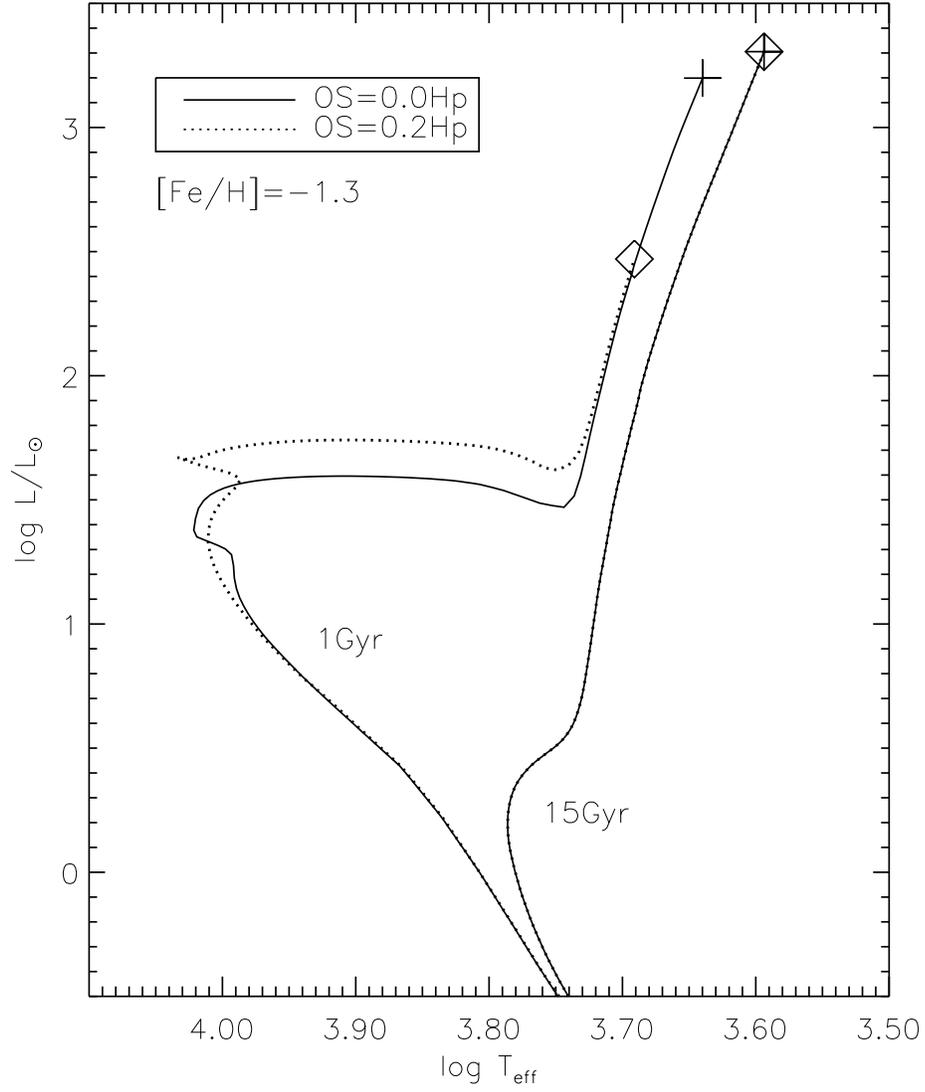}
\caption{
The effects of convective core overshoot (OS) parameter to the 
shape of isochrones. An increase in the adopted amount of OS causes
a brighter MSTO and a longer lifetime in the hydrogen burning phase.
The RGB tips are marked with $diamonds$ (with OS) and $plus$ signs 
(without OS).
}
\end{figure}

\begin{figure}
\plotone{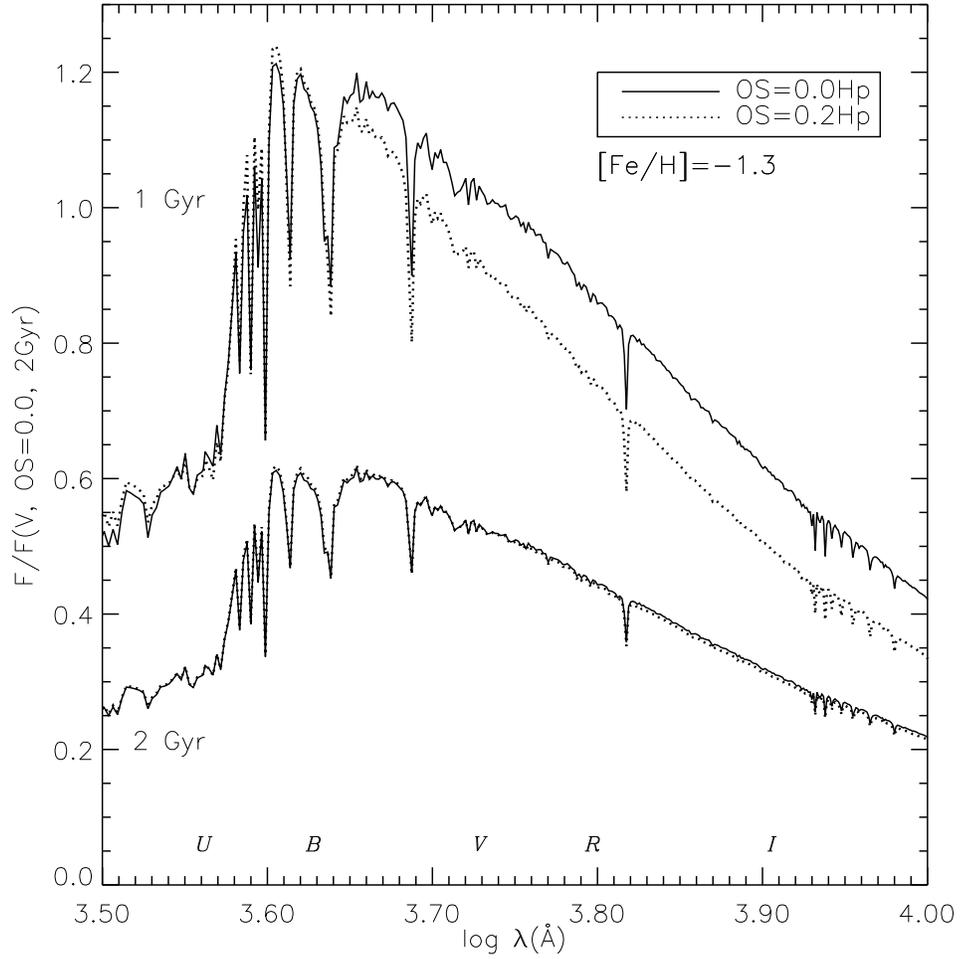}
\caption{
The effects of OS to the integrated spectrum. A larger value of OS parameter 
causes the population models to appear bluer, but only at small ages.
In order to single out this effect, all other input parameters are kept 
the same as in our standard model (see Table 1).}
\end{figure}

\begin{figure}
\plotone{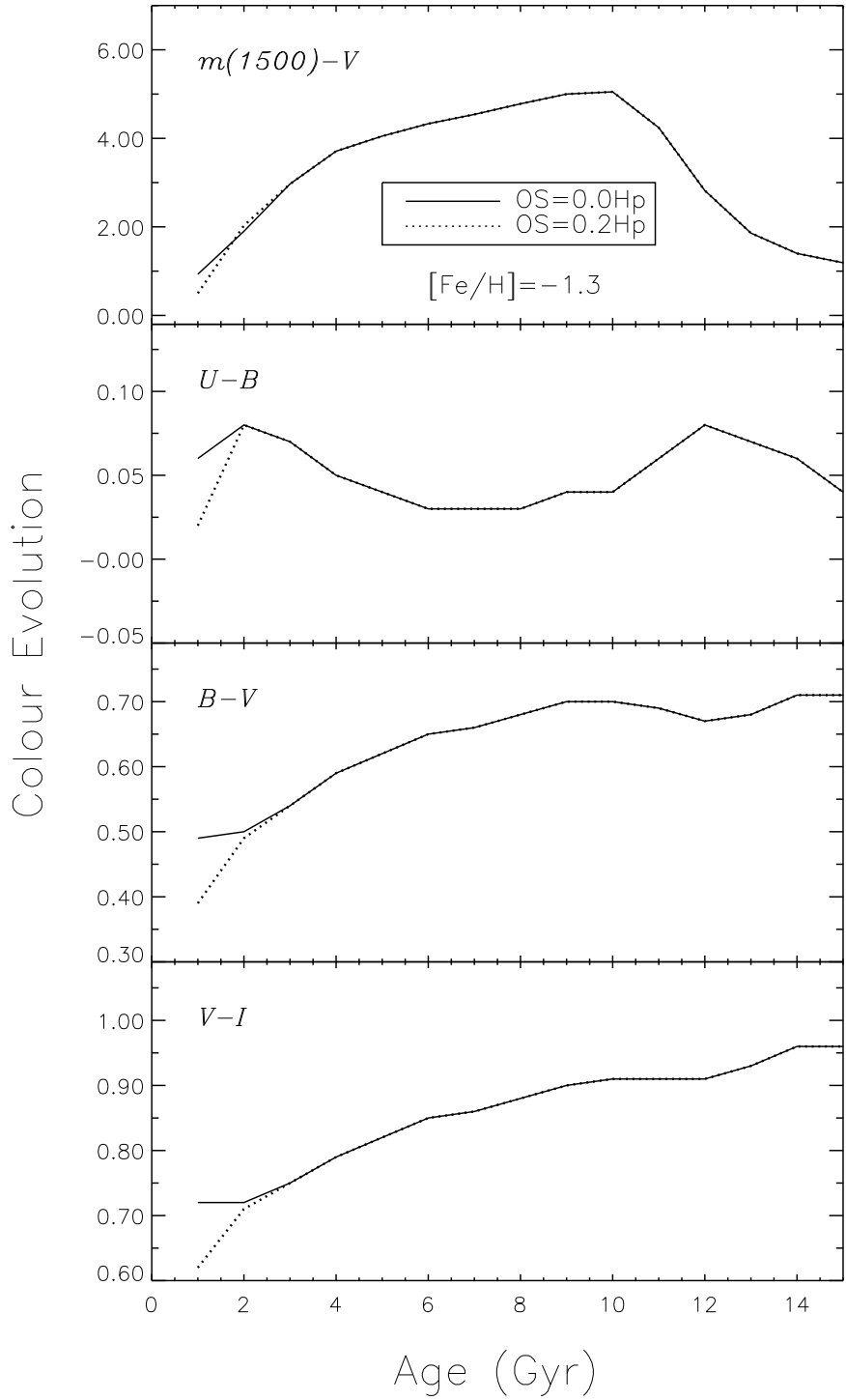}
\caption{The effects of OS to the integrated colors. A larger value of 
OS parameter causes the population models to appear bluer at small ages.
}
\end{figure}

\begin{figure}
\plotone{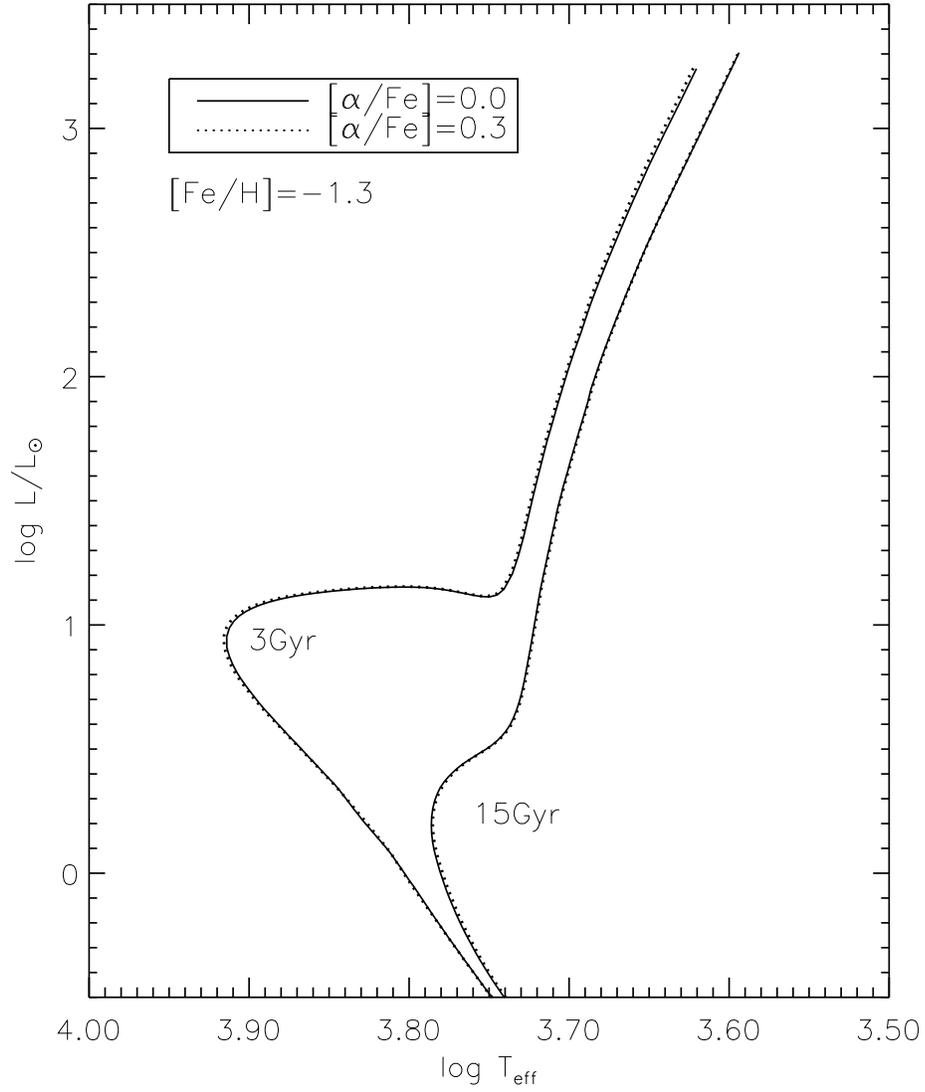}
\caption{The effects of $\alpha$-enhancement to the 
shape of isochrones. When two low-metallicity models with the same total 
metal abundance ($Z$) are compared, $\alpha$-enhancement does not cause any 
noticeable difference in stellar evolution models, and thus in color 
evolution models. }
\end{figure}

\begin{figure}
\plotone{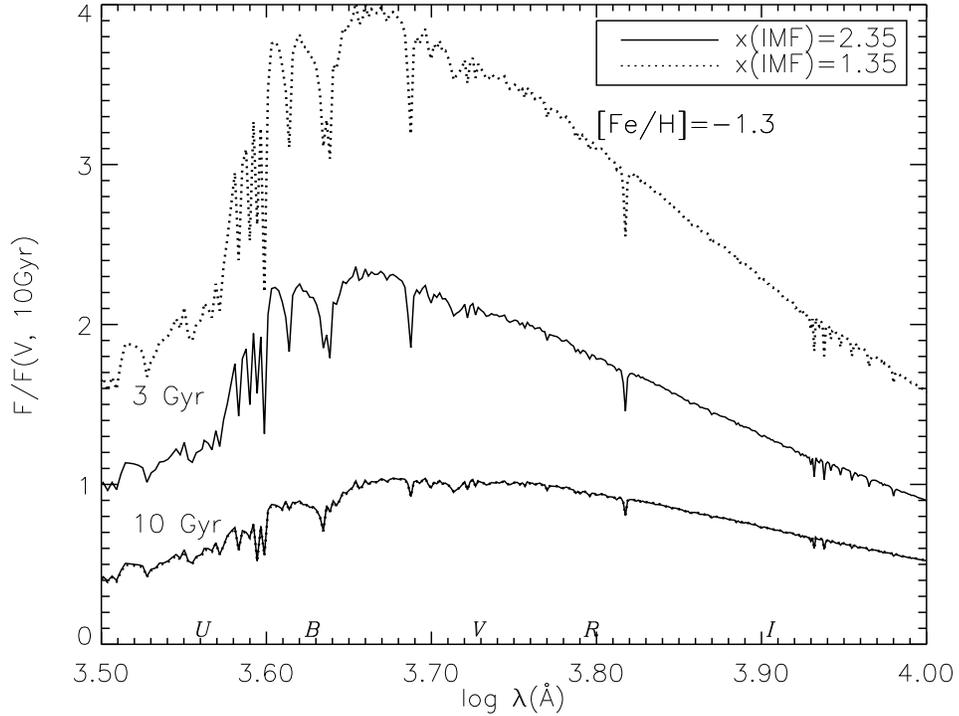}
\caption{The effects of the IMF slope to the integrated spectrum. 
The younger models are normalized to their older (10\,Gyr) counterpart.
Because the normalisation has been made by the luminosity of the old model, 
these two populations have substantially different total masses, 
which is why the young model with x(IMF)=1.35 (top dotted model) has 
a significantly higher flux level than its counterpart
with x(IMF)=2.35. Note the two old models, although with different IMF slopes,
are indistinguishable.
In order to single out this effect, all other input parameters are kept 
the same as in our standard model (see Table 1).}
\end{figure}

\begin{figure}
\plotone{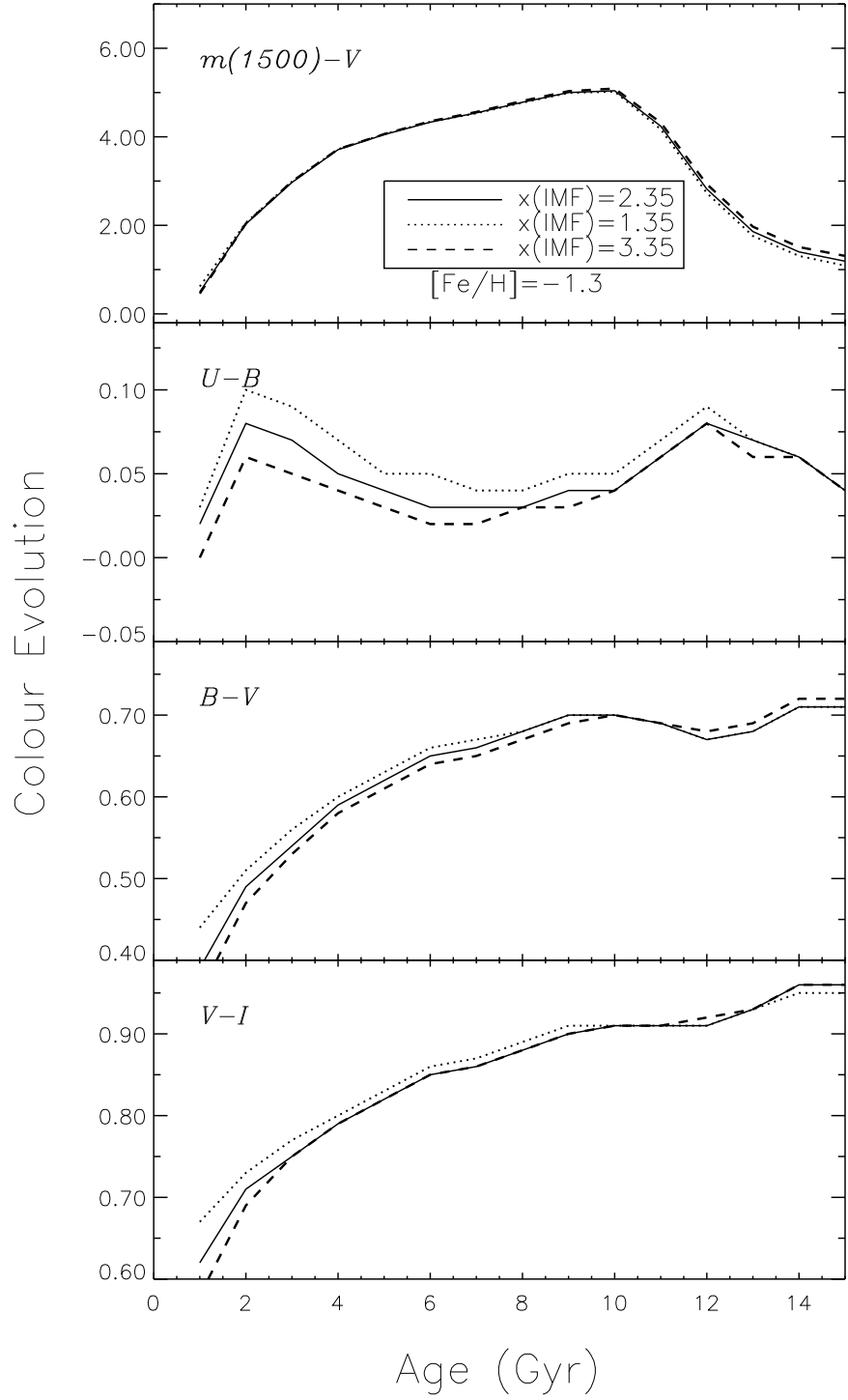}
\caption{The effects of the IMF slope to the integrated colors. }
\end{figure}

\begin{figure}
\plotone{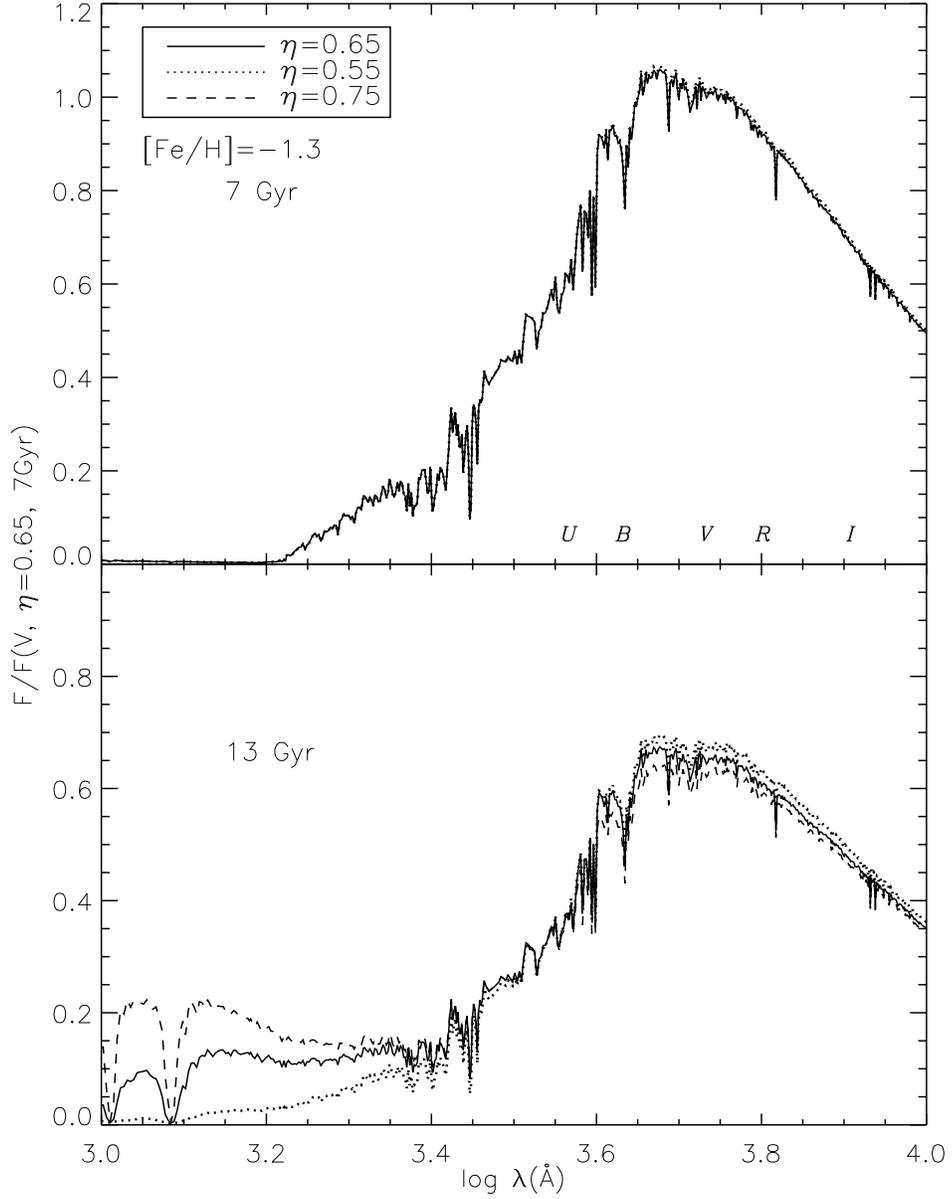}
\caption{The effects of the mass loss efficiency parameter ($\eta$) to the 
integrated spectrum. 
In order to single out this effect, all other input parameters are kept 
the same as in our standard model (see Table 1).}
\end{figure}

\begin{figure}
\plotone{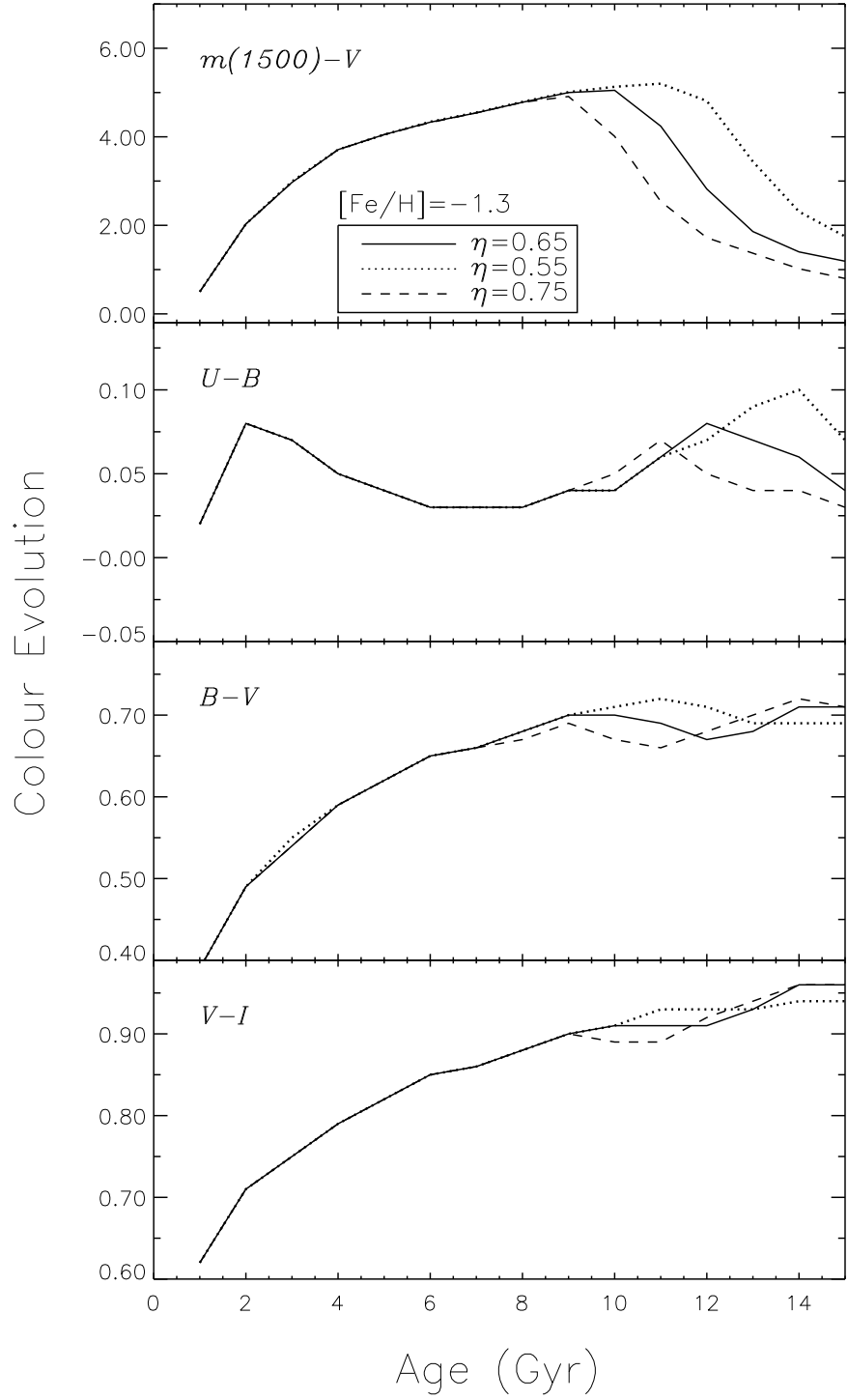}
\caption{The effects of the mass loss efficiency parameter to the 
integrated colors.}
\end{figure}

\begin{figure}
\plotone{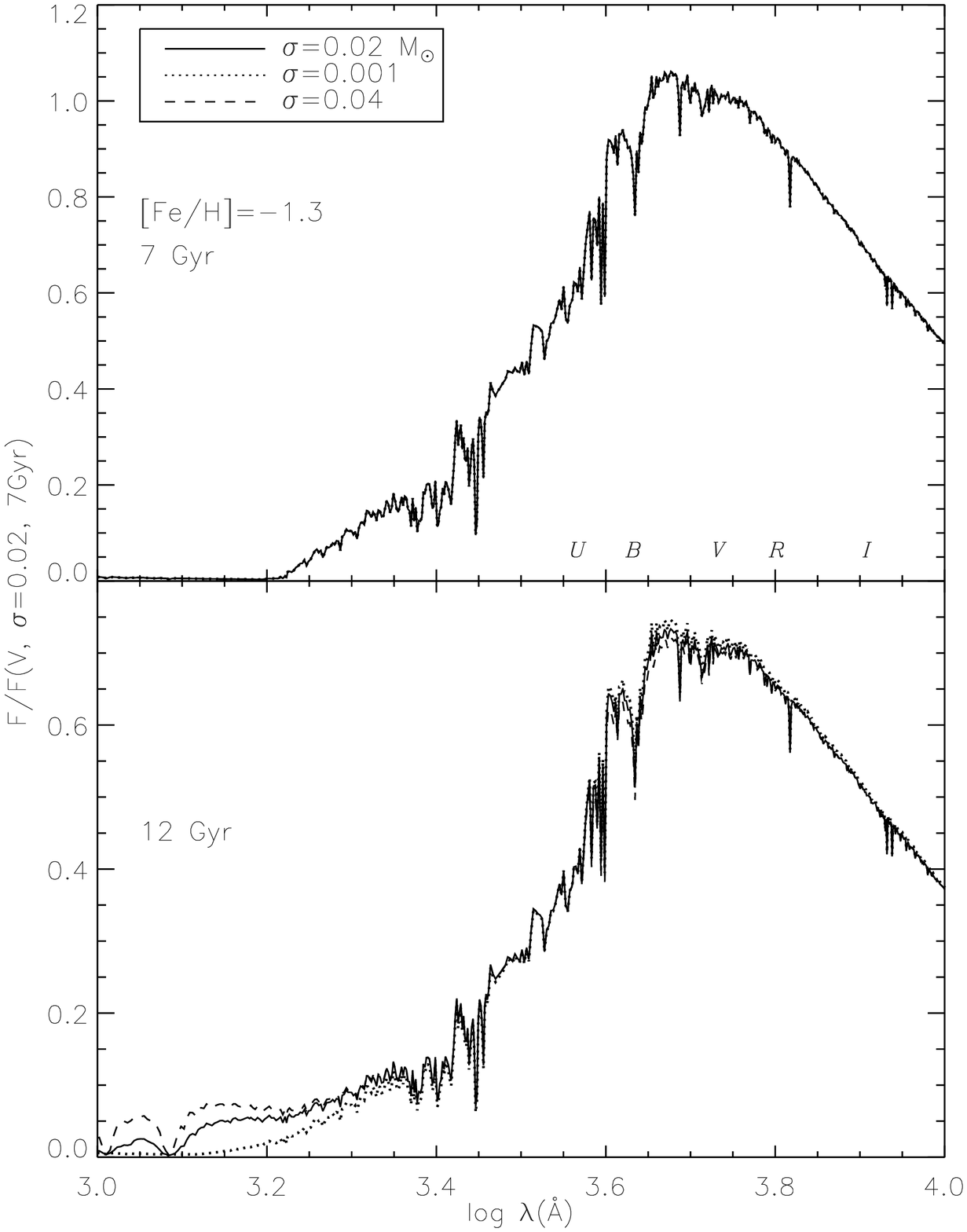}
\caption{The effects of the mass loss dispersion parameter (\sigHB) to the 
integrated spectrum. 
In order to single out this effect, all other input parameters are kept 
the same as in our standard model (see Table 1).}
\end{figure}

\begin{figure}
\plotone{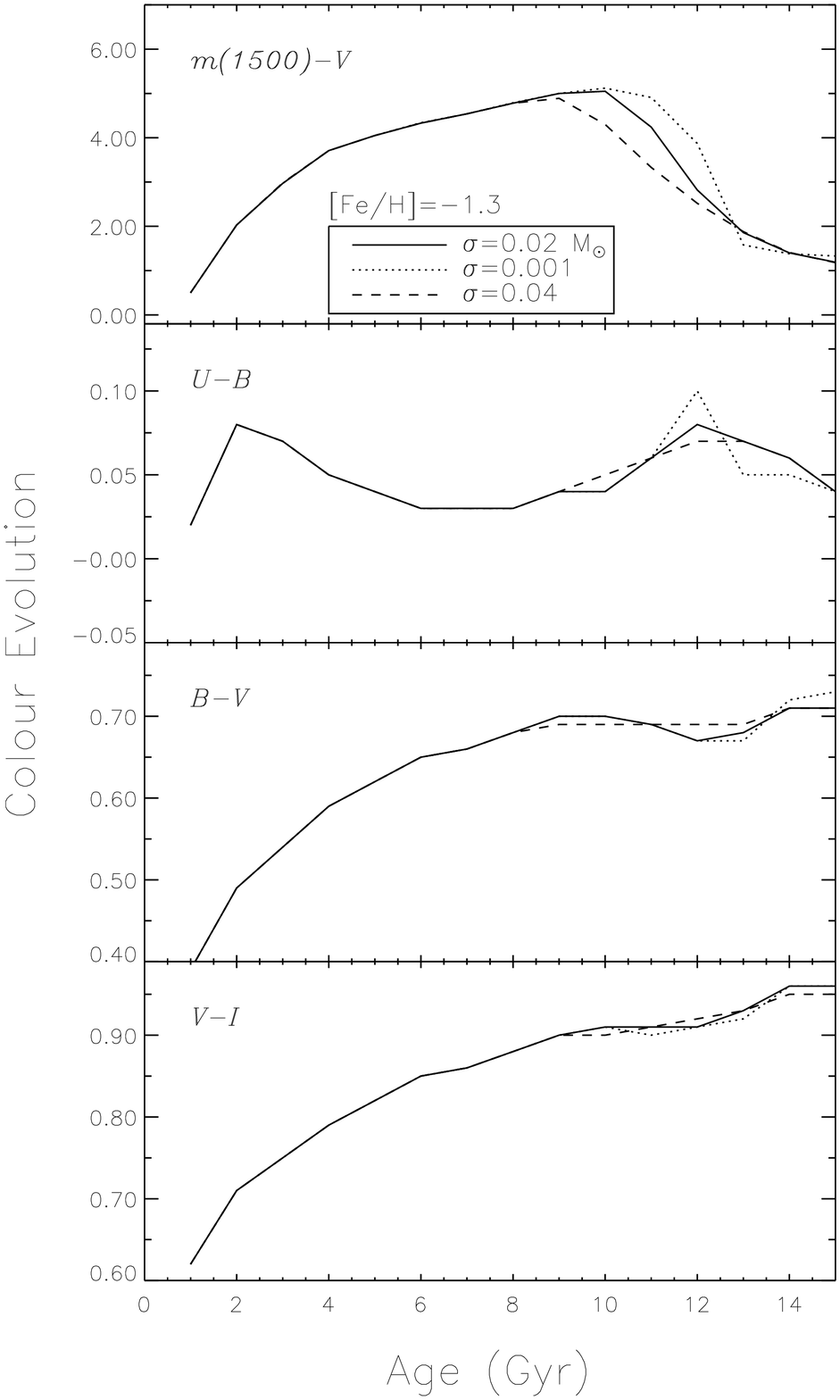}
\caption{The effects of the mass loss dispersion parameter (\sigHB) to the 
integrated colors.}
\end{figure}

\begin{figure}
\plotone{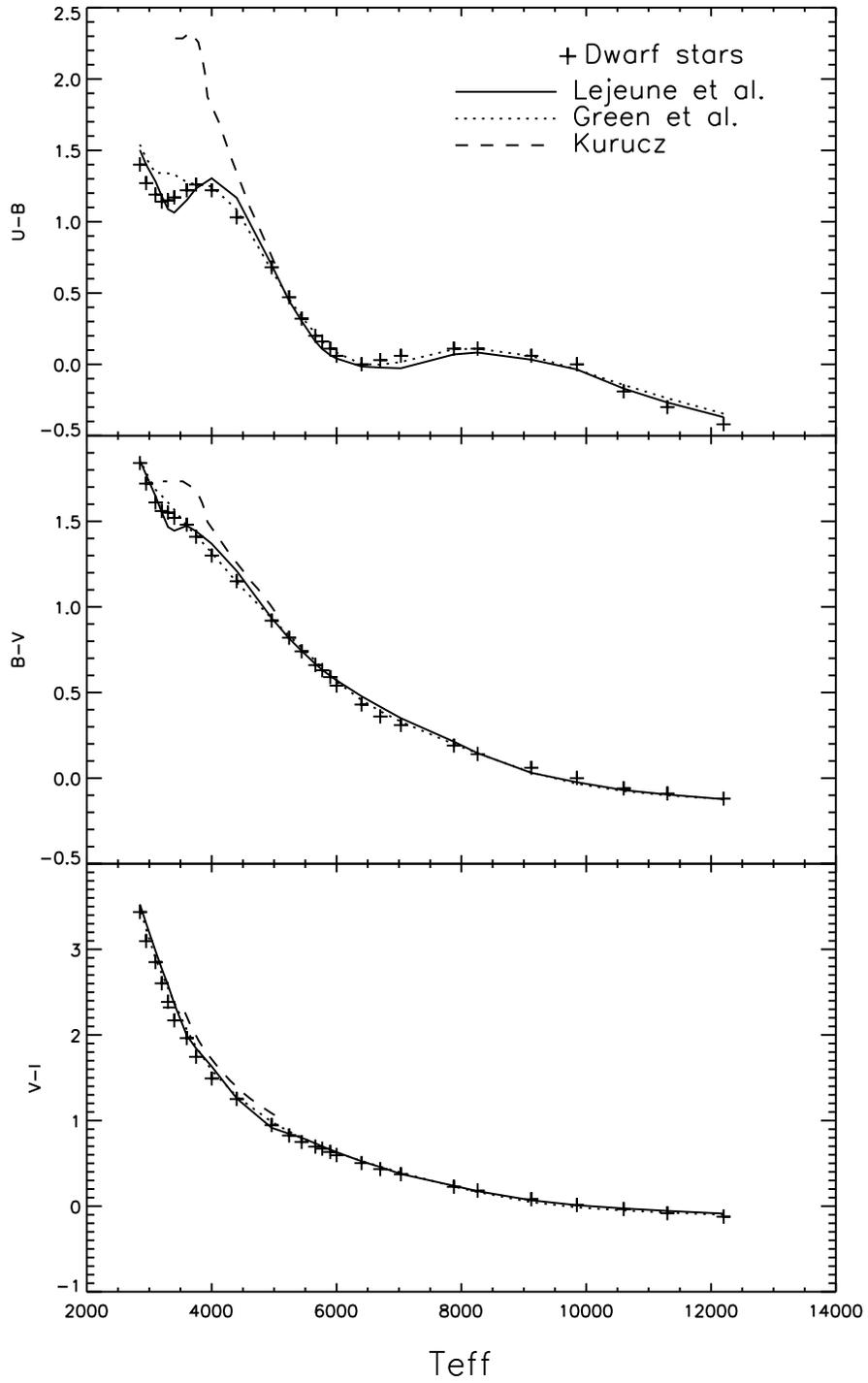}
\caption{The difference between various color transformation tables for
the solar metallicity main-sequence stars.
Continuous lines show the LCBK color transformation table, dotted lines are
for the Green et al. table (GDK), and dashed lines are for 
the Kurucz table. Empirical stellar data have been kindly provided by
Guy Worthey.}
\end{figure}

\begin{figure}
\plotone{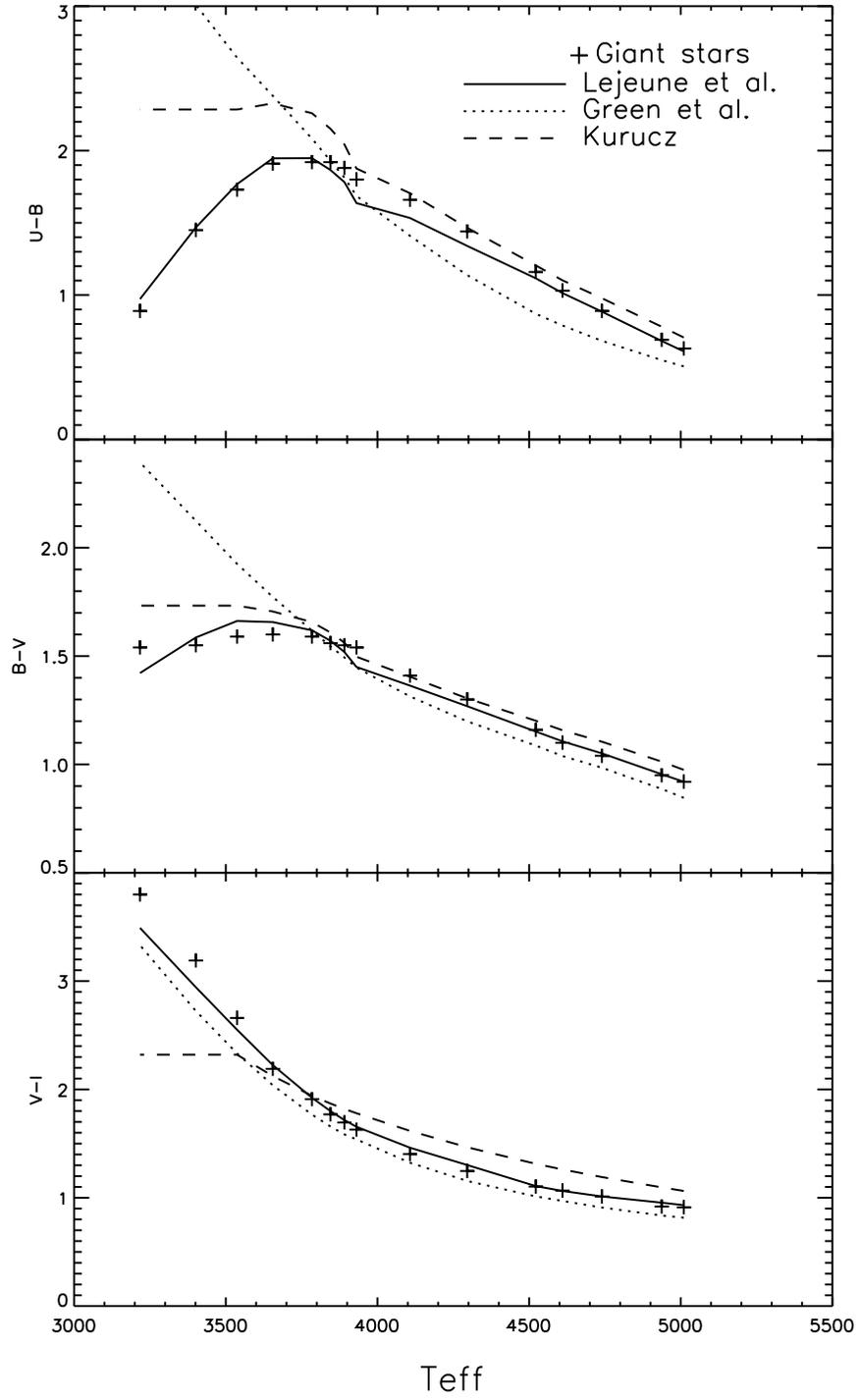}
\caption{Same as Figure 14, but for red giants.}
\end{figure}

\begin{figure}
\plotone{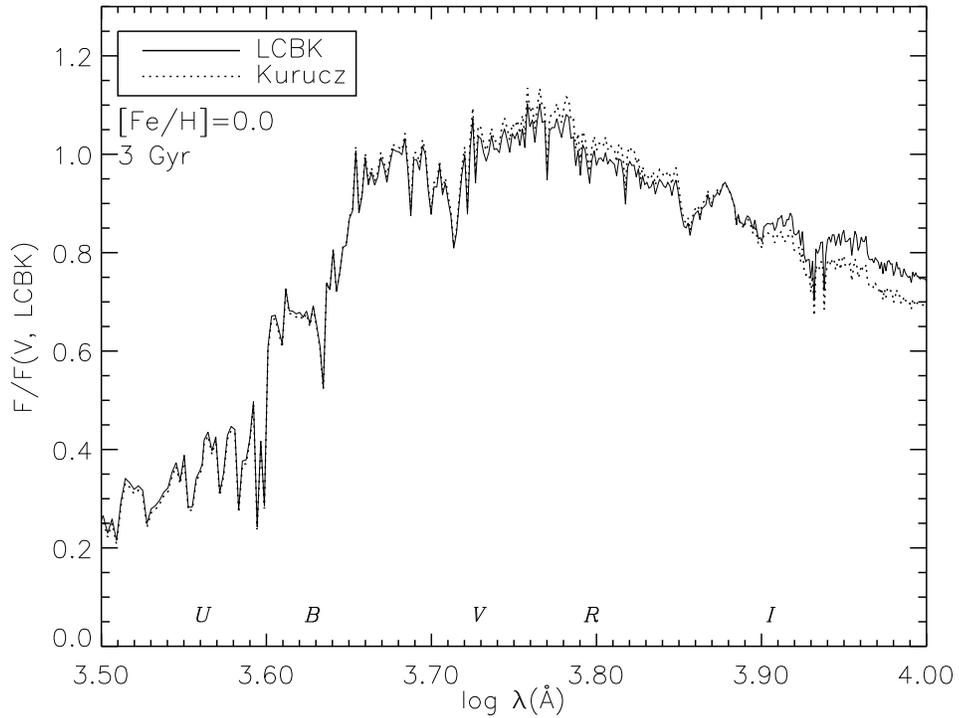}
\caption{
The effects of the use of different spectral libraries to the 
integrated spectrum. When all other input parameters are fixed, 
the use of LCBK library, in comparison to the Kurucz library, 
causes the model spectrum appear redder.
In order to single out this effect, all other input parameters are kept 
the same as in our standard model (see Table 1).}
\end{figure}

\clearpage

\begin{figure}
\plotone{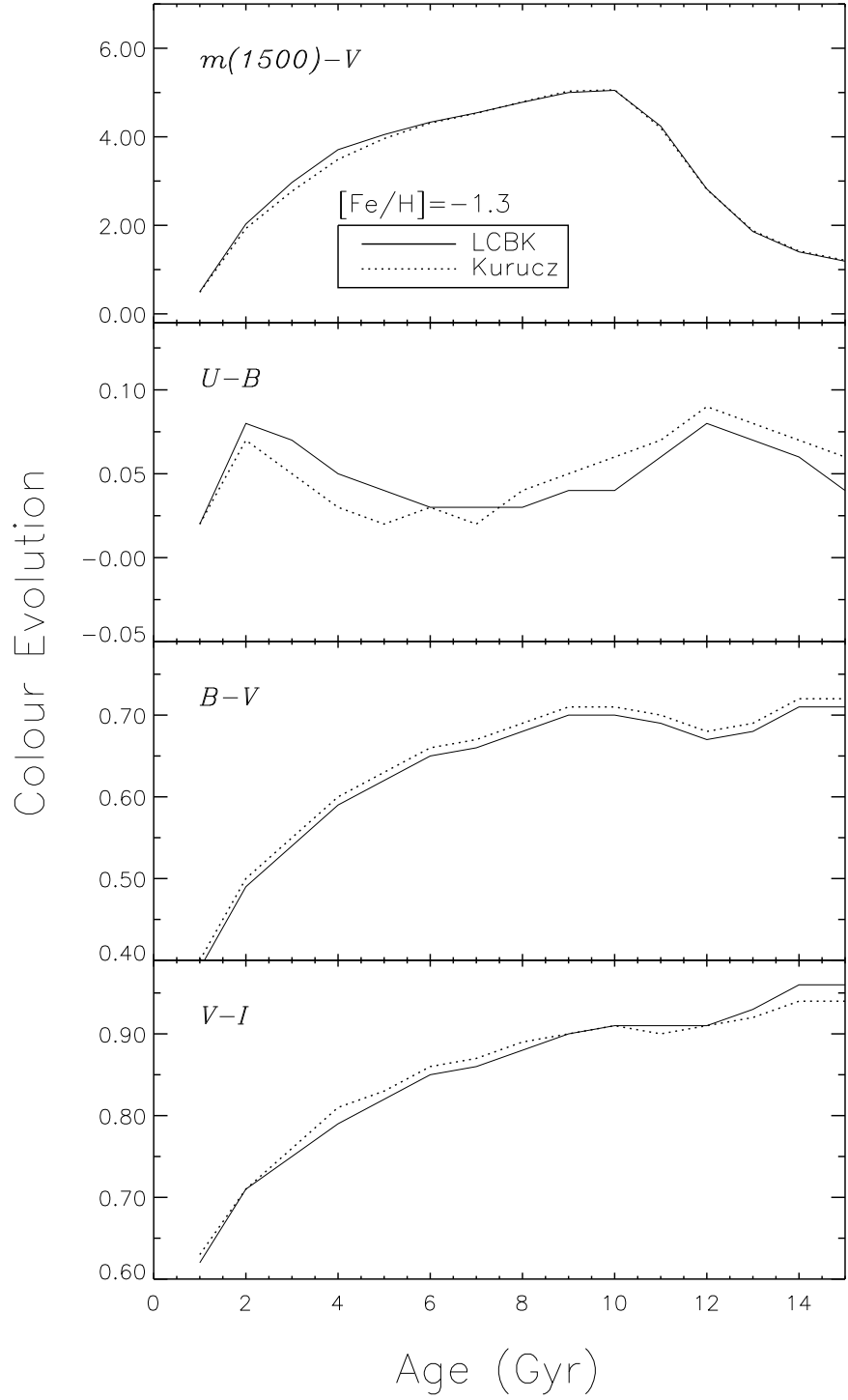}
\caption{The effects of the use of different spectral libraries to integrated
colors.}
\end{figure}

\begin{figure}
\plotone{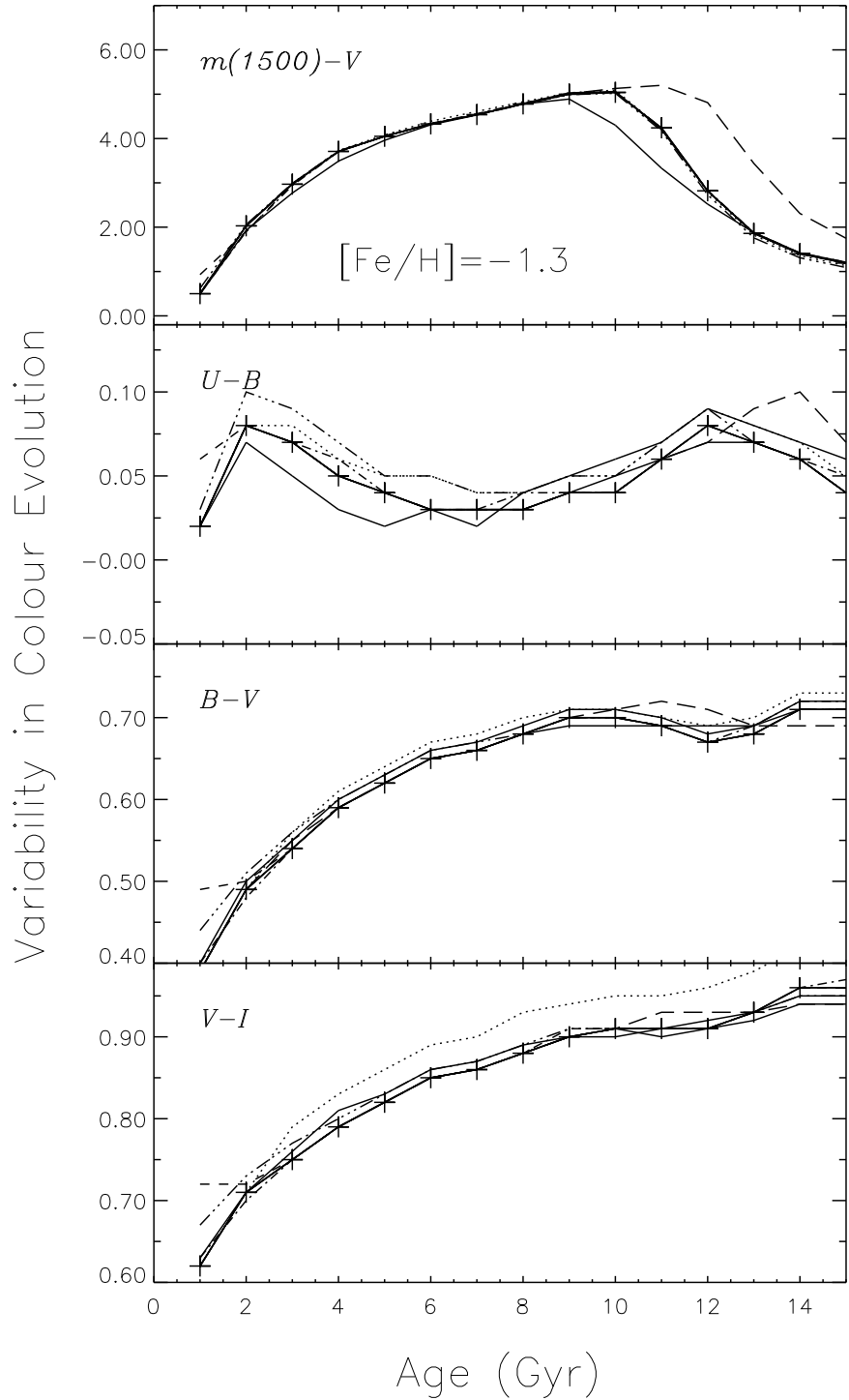}
\caption{The effects on integrated colors from all uncertain input 
parameters investigated in this study. Legends are purposely omitted 
in this plot,
as we want to just demonstrate the overall level of uncertainty coming from 
different assumptions. Our standard models are denoted by connected $plus$ 
symbols.}
\end{figure}

\begin{figure}
\plotone{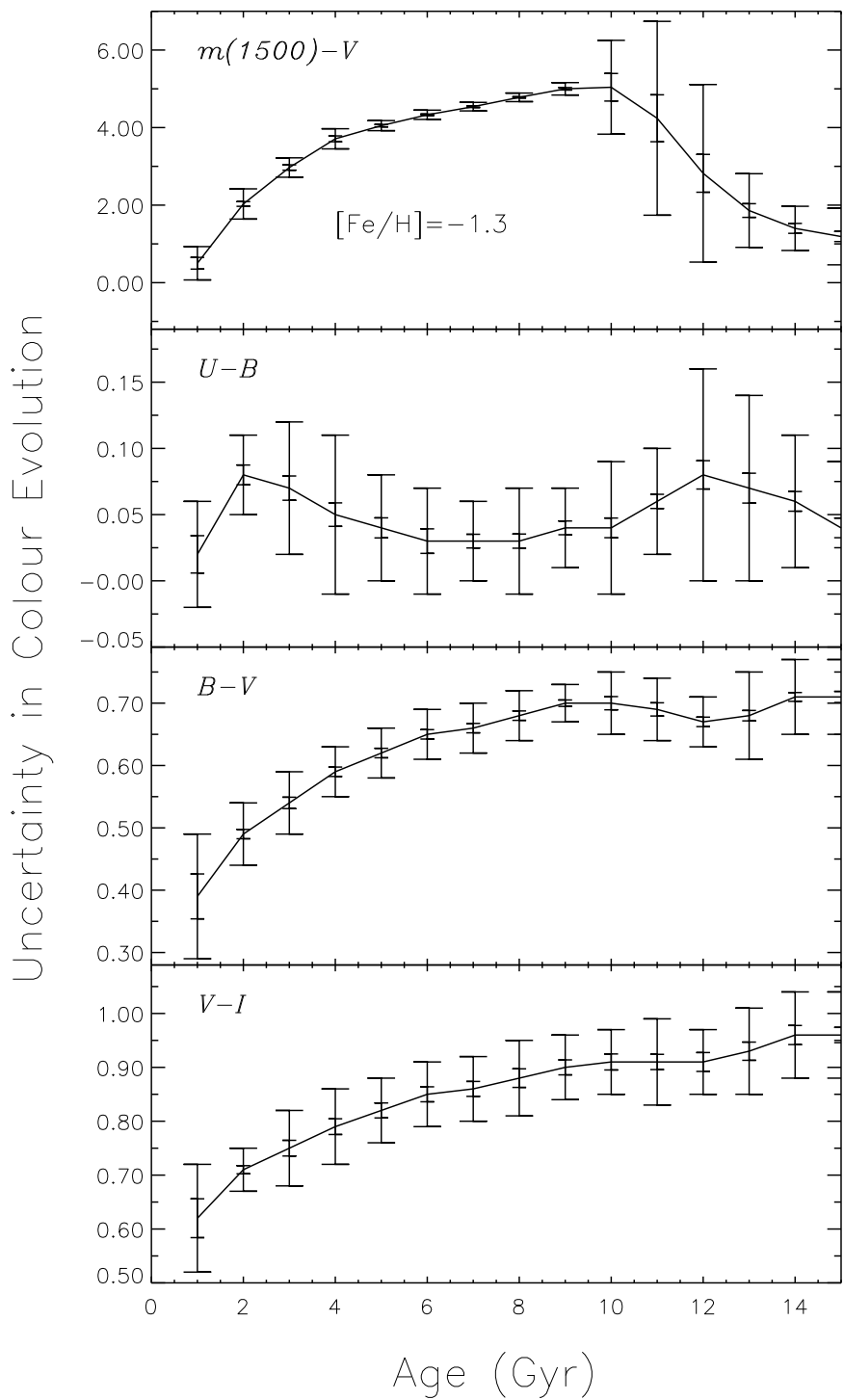}
\caption{The combined total uncertainty in color evolution originating
from all sources investigated in this study. The inner (smaller) error bars 
are the standard deviation and the outer (larger) bars are the maximum 
uncertainty in case all uncertainties build up in the same direction.
The mean curve is our standard model.
}
\end{figure}

\begin{figure}
\plotone{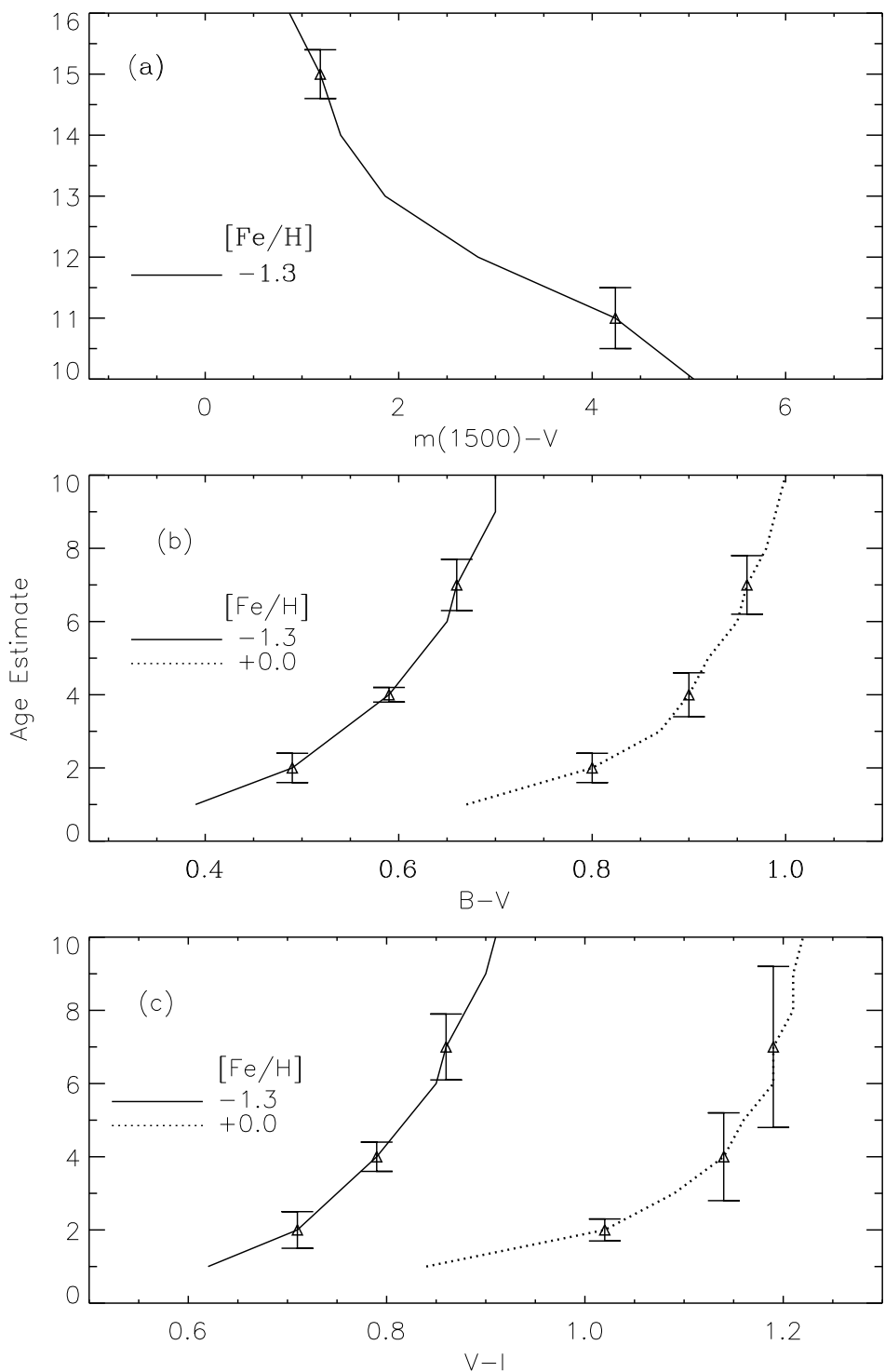}
\caption{Uncertainty in age estimates caused by the uncertainties in
various input physics. Continuous lines are metal-poor models, and
dotted lines are metal-rich ones. We believe that metal-rich models 
are less accurate because mass loss in metal-rich stars is not sufficiently
well constrained yet. For this reason, they are omitted in (a). 
At small ages, optical colors are better age indicators, while at
large ages \uvv is better. The uncertainties in age estimates are
small enough that such age estimates are useful.}
\end{figure}

\begin{figure}
\plotone{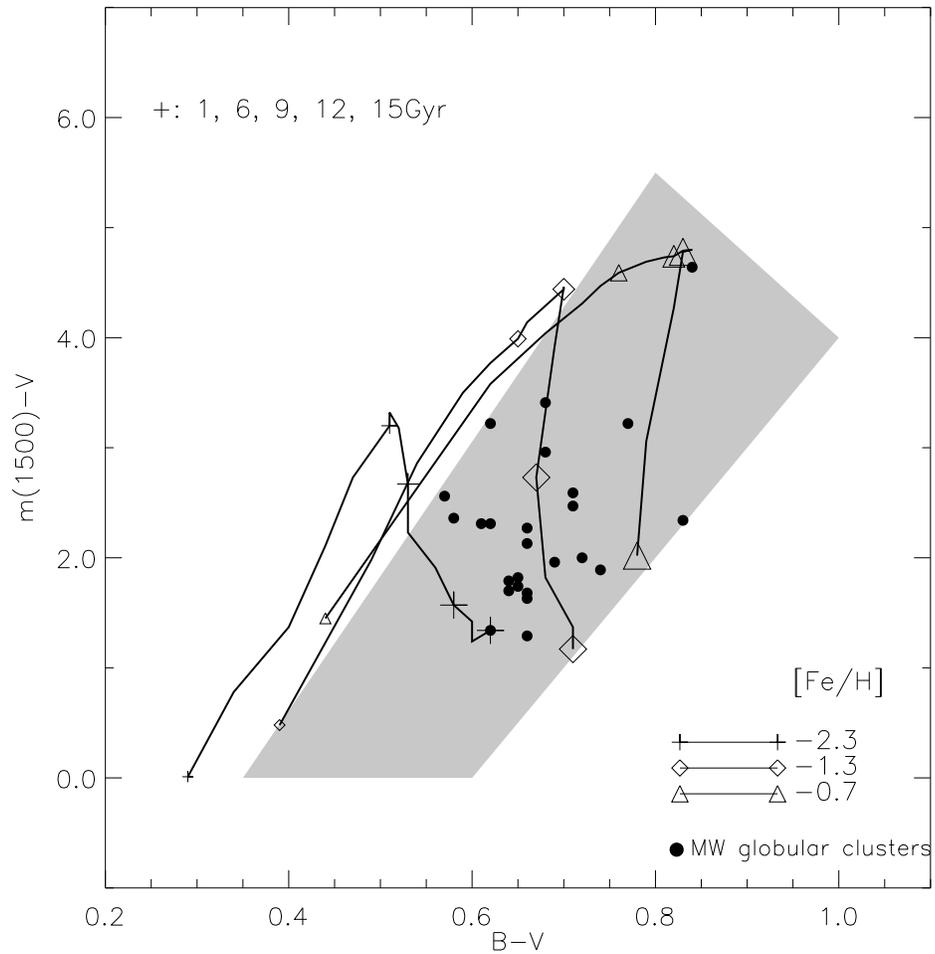}
\caption{Color-color diagram for metal-poor simple stellar populations.
Three sequences are for different metallicities. Each sequence is an
age sequence covering the ages of 1 -- 15 Gyr. From the smallest symbol
to the largest, five age points in increasing size (1, 6, 9, 12, \& 15\,Gyr) 
are specially marked. Old populations should appear in the shaded region.
The Galactic globular clusters are all in the box as expected from their
large ages based on isochrone fitting. The data are from Dorman et al. (1995)
and Harris (1996).
}
\end{figure}

\end{document}